ORIGINAL ARTICLE

# Security and privacy in virtual reality: a literature survey

**Alberto Giaretta**[1]

Received: 9 February 2024 / Accepted: 30 November 2024

**Abstract**
Virtual reality (VR) is a multibillionaire market that keeps growing, year after year. As VR is becoming prevalent in households and small businesses, it is critical to address the effects that this technology might have on the privacy and security of its users. In this paper, we explore the state-of-the-art in VR privacy and security, we categorise potential issues and threats, and we analyse causes and effects of the identified threats. Besides, we focus on the research previously conducted in the field of authentication in VR, as it stands as the most investigated area in the topic. We also provide an overview of other interesting uses of VR in the field of cybersecurity, such as the use of VR to teach cybersecurity or evaluate the usability of security solutions.

**Keywords** Virtual reality (VR) · Privacy · Security · Cybersecurity · Authentication

## 1 Introduction

Extended reality (XR) is a paradigm that refers to different types of human–machine interaction within real and virtual environments, enabled by computer technology and wearable devices. Under the umbrella of XR, Virtual Reality (VR) is a computer-generated simulation of a 3D environment that enables realistic physical interactions by means of technology and wearable devices. The popularity of VR has grown steadily over the years; according to Oberlo (2022), the number of US consumers that used VR went from 16% in 2019 to 19% in 2020. Of those users, shows a report from Petrock (2020), 52.1 million people use VR technology at least once per month. Furthermore, the growth is expected to continue in the near future. By 2028, the compound annual growth rate of the VR market is expected to reach 18% (Research 2022), according to statistics by Grand View Research. In terms of market value, Valuates Reports (Reports 2022) projects that the global VR market size will reach US$ 26,860 million by 2027, from US$ 7719.6 million in 2020.

By design, VR is capable of collecting a large amount of non-verbal information, such as users' movements, biometrics, and usage patterns. As shown in Fig. 1, VR devices are equipped with different sensors that collect both explicit input and non-verbal information, which is used by the VR engine to model the virtual world according to the user's actions. As noted by Bailenson (2018), non-verbal information is uniquely telling and can be used for different goals, from tailoring advertisements, to determining if users are low or high performers. Hours of personal use, paired with unclear policies regarding data handling and learning algorithms, could allow companies to estimate users' preferences and infer their characteristic behaviour. Taking into account the consistent and considerable growth of VR, alongside its pervasive nature, it is of paramount importance to evaluate any privacy and security concern that could arise. In this paper, we provide a literature survey on VR threats and issues, in terms of privacy and security, with the aim of highlighting what has been done and what is still open for research.

### 1.1 Related works and contributions

Within the umbrella term XR, besides VR, exist other types of extended reality. Namely, Augmented Reality (AR) and Mixed Reality (MR). In AR, virtual objects are added to the real environment for enriching it;a notable example is pointing a smartphone to an art piece and getting information about the piece itself. MR is a combination of VR and AR, where the interactions do not happen exclusively in the virtual space nor in the real one, but in a hybrid fashion. AR and MR exhibit their own set of challenges and issues, in terms of security and privacy, which are not necessarily similar to those found in VR. In 2014, Roesner et al. (2014) conducted a survey on AR security and privacy concerns, while King et al. (2020) focused specifically on the privacy issues of AR applications. Regarding MR, De Guzman et al. (2019) conducted an exhaustive survey on the security and privacy issues that affect it.

✉ Alberto Giaretta
alberto.giaretta@oru.se

[1] Örebro University, Fakultetsgatan 1, 70218 Örebro, Sweden

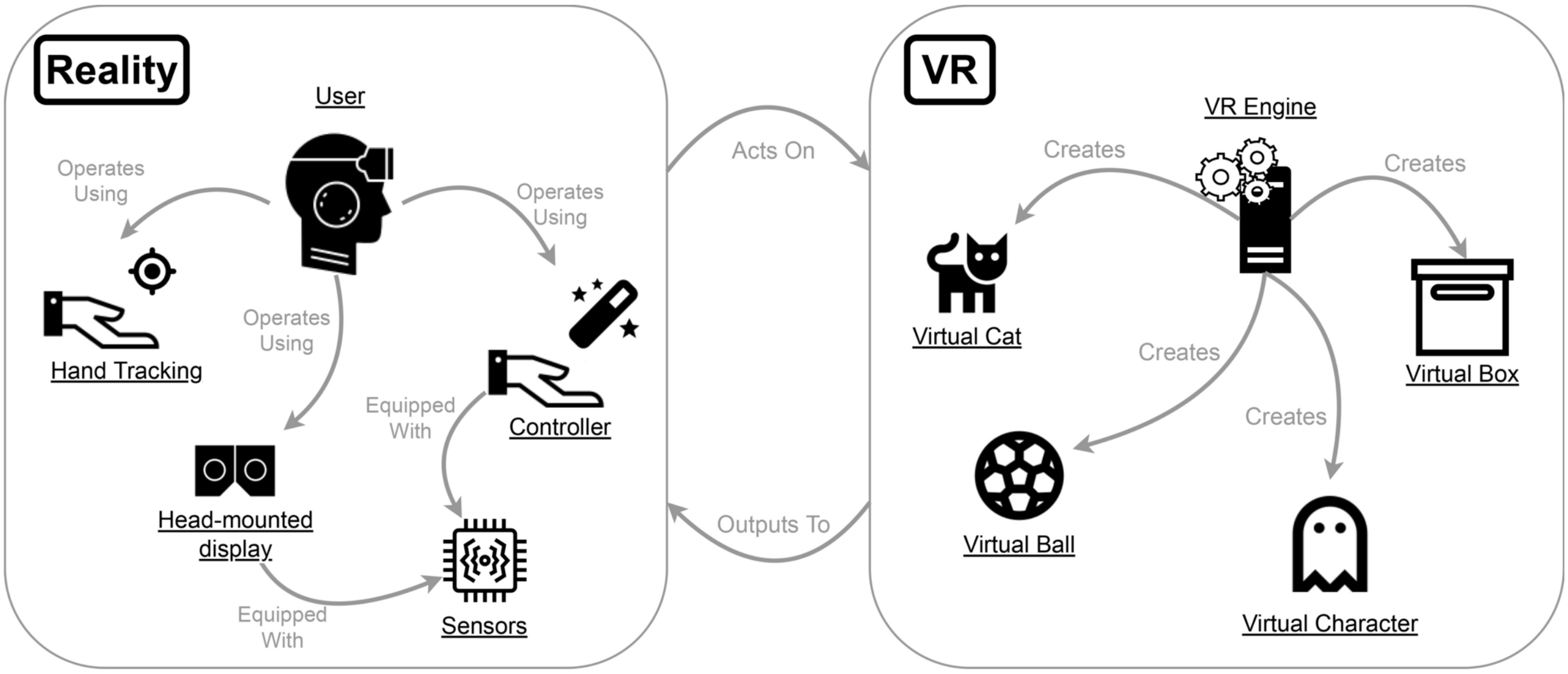

**Fig. 1** The Virtual Reality (VR) paradigm. Users perceive the VR environment using head-mounted displays (HMDs) and interact with it via controllers or more advanced techniques, such as finger tracking sensors. Every device is equipped with various sensors to provide useful information to the VR engine, which creates the virtual world and modifies it according to the input received from the user

On the topic of VR security and privacy, a few works have been published in the past. Stephenson et al. (2022) in 2022 conducted a large systematisation of knowledge (SoK) study on authentication mechanisms in AR/VR environments. Although a considerable percentage of the papers published on the topic of VR security are concerned with authentication and identification issues, as we also highlight in our paper, that is only one aspect of the matter at hand. In this paper, we present other papers that are not related to authentication mechanisms, but that are still extremely important for gaining a complete perspective on the topic (such as works that propose new side-channel attacks, privacy aspects, or works that use VR for improving cybersecurity awareness). Other works on authentication in VR exist, such as the survey by Heruatmadja et al. (2023) on the use of biometrics for authentication purposes. However, as previously said, these works cover only a subset of the aspects covered in this survey.

Odeleye et al. (2023) in 2022 conducted a survey on cybersecurity and privacy in VR, with a detailed categorisation of cybersecurity threats and challenges. Differently to what we have done in this paper, the authors treat privacy as one of the challenges stemming from security issues. On the one hand, this is helpful for the reader to delve deep into the effects of cybersecurity attacks on privacy. On the other hand, their survey does not dedicate space to reasoning upon effects and consequences on VR users, in case of failed privacy assurances. Cayir et al. (2024) published a SoK on this topic as well. Although quite updated with a recent perspective on the topic, the paper is not an in-depth survey (as stated by the authors themselves) but a SoK that forced the authors to carefully select only the most interesting papers for their scope, resulting in a selection of 41 papers. In the paper, the authors produce a taxonomy that is similar but less detailed than the taxonomies we introduced in this paper.

Unlike all the previous manuscripts we have discussed so far, our paper not only discusses authentication mechanisms in VR or privacy and cybersecurity threats. We also provide a new perspective on the use of VR technology to improve and strengthen cybersecurity awareness and expertise. For example, we summarise various works that use VR for teaching cybersecurity to undergrad students. We also cover a number of papers that use VR for training physical and cyber-physical security, and papers that show how VR can be used as a tool for evaluating the usability of new cybersecurity mechanisms. It is also worth noting that, apart from the survey by Cayir et al. (2024), the other surveys mentioned above have been published prior to 2023. As VR is a fast-moving technology that has been evolving quite rapidly throughout the last years, it is important to maintain an updated overview of the state-of-the-art. To summarise, in this paper we provide four main contributions:

- A taxonomic overview of privacy issues that affect VR, organised by typology and causes;
- A taxonomy of threats to VR security, categorised in traditional attacks that also affect VR, such as denial of service (DoS), and attacks specific for VR, such as dizziness-inducing ones;
- An updated and in-depth exploration on the research done on authentication aspects in VR;
- An overview of other uses of VR in the field of cybersecurity and physical security, such as teaching cybersecurity, training employees, or evaluating the usability of new cybersecurity solutions.

### 1.2 Outline

This paper is organised as follows. In Sect. 2 we describe the methodology followed for conducting this literature survey. In Sect. 3, we discuss privacy issues and threats within the VR domain. Similarly, in Sect. 4 we discuss the threats to the security of VR, and we dedicate Sect. 5 to the topic of authentication for VR. In Sect. 6 we discuss different uses of VR within the cybersecurity domain, such as teaching security concepts, training physical security, and evaluating the usability of security processes. Finally, in Sect. 7 we provide some final remarks and provide our conclusions.

## 2 Research methodology

In this section, we describe the research methodology used in this paper to investigate the different aspects of security and privacy related to VR. The main goal of this survey is not to provide a complete systematic review of the research field. However, our methodology takes inspiration from the guidelines provided by Petersen et al. (2015) for conducting systematic mapping studies.

### 2.1 Research questions

In this paper, we aim to examine the extent and nature of the research activity for what it concerns privacy and security within the VR realm. Therefore, we have identified two main research questions:

- **RQ1** What are the privacy issues that affect VR?
- **RQ2** What are the security threats to VR?

During our initial investigation, we have encountered unexpected results when assessing the extent of research in some topics. In particular, we found that the research community has focused a lot on investigating user authentication methods in VR. Therefore, we have expanded the first two research question with an additional one:

- **RQ3** What is the state-of-the-art in VR user authentication?

We also have encountered surprising results regarding the nature of some research. We have detected a number of research papers that investigate the use of VR to improve cybersecurity training and teaching, as well as the use of VR to evaluate security solutions. This led us to adding a fourth and last research question:

- **RQ4** Which security research areas can benefit from the use of VR?

### 2.2 Research databases, keywords, and queries

For the scope of this study, we have restricted our search to four main publishers and their research databases, namely Springer, IEEE, ACM, and Elsevier. In addition, we have conducted a manual search in the USENIX database, as it is the main independent conference in the field of cybersecurity, and it is not indexed by the other four databases.

Following the research questions previously mentioned in Sect. 2.1, we have identified a set of relevant keywords to conduct the first research phase. For investigating RQ1, we have selected the query *"virtual reality" AND "privacy"*. For RQ2 and RQ4, we searched for *"virtual reality" AND ("security" OR "cybersecurity")* This is due to the fact that, in the field of cybersecurity, some researchers refer to the field as *security* while others prefer the other term. Last, to investigate in detail the authentication aspects, we have used the query *"virtual reality" AND ("authentication" OR "identification" OR "profiling")*. For the sake of brevity, we omit the specific queries for each research database, as they exhibit slight different grammar rules.

A first search of these terms over the entire text and metadata of the manuscripts, resulted in many papers for each database. The aggregated results of the terms in the four databases totalled up to more than 50,000 papers, mainly due to the poor filtering performance of some research portals.

### 2.3 Inclusions and exclusion criteria

Based on the vast number of results obtained in the first search, we have formulated inclusion and exclusion criteria. Regarding the inclusion criteria, when possible (e.g., in IEEE Xplore), we have restricted our research to the text found in the *manuscript title field*, the *keywords* field, and the *abstract* field. This allowed us to exclude most papers where the occurrence of search words is not intentional, but merely incidental. Furthermore, we have included only research articles, in particular conference papers and journal papers. We excluded posters and extended abstracts due to their limited contributions, although we included some short manuscripts when we deemed their approach substantially interesting. Out of scope were also papers dealing with Augmented Reality (AR), Mixed Reality (MR), or Extended Reality (XR) in general. In addition, we have decided to exclude the body of literature focused on metaverse. Although the metaverse is correlated with VR technology, the set of privacy and security challenges that characterise them are different. For example, papers that discuss security and privacy in the metaverse are concerned with threats that arise from the interaction between virtual avatars, or with the conjunction of different virtual environments in a larger virtual world.

Once applied this first filtering, we were left with a total number of 33,000 papers, approximately. After an assessment of the quality of the results, we realised that a lot of the retrieved manuscripts were outside the scope of this paper, regardless of the aforementioned filtering steps. For example, the search returned various papers where the authors used the term security, but in a different context (e.g., security guards), or as a passing comment, or for referring to military defence. For example, searching for *"virtual reality" AND "security" OR "cybersecurity"* among SpringerLink articles, returns as a first-page result a paper concerned with the use of VR for diagnosing ADHD, clearly out of the scope of our search. Therefore, we scanned all the titles and abstracts of the results to the best of our capabilities, given the large number of papers. This process made us select a final pool of 146 relevant papers. Upon further and detailed inspection, 17 of the 142 papers did not fit the criteria defined for the scope of this survey. After excluding also these papers, we were left with a final pool of 125 works to review in this survey.

## 3 Taxonomy of privacy issues in VR

VR technology raises a number of concerns about the privacy of its users. In this section, we provide an overview of the different threats to privacy in the context of VR, following the structure shown in Fig. 2. In 2018, Adams et al. (2018) identified to which extent users and developers are concerned about privacy in VR. They conducted interviews with 20 participants, 10 users and 10 developers, showing

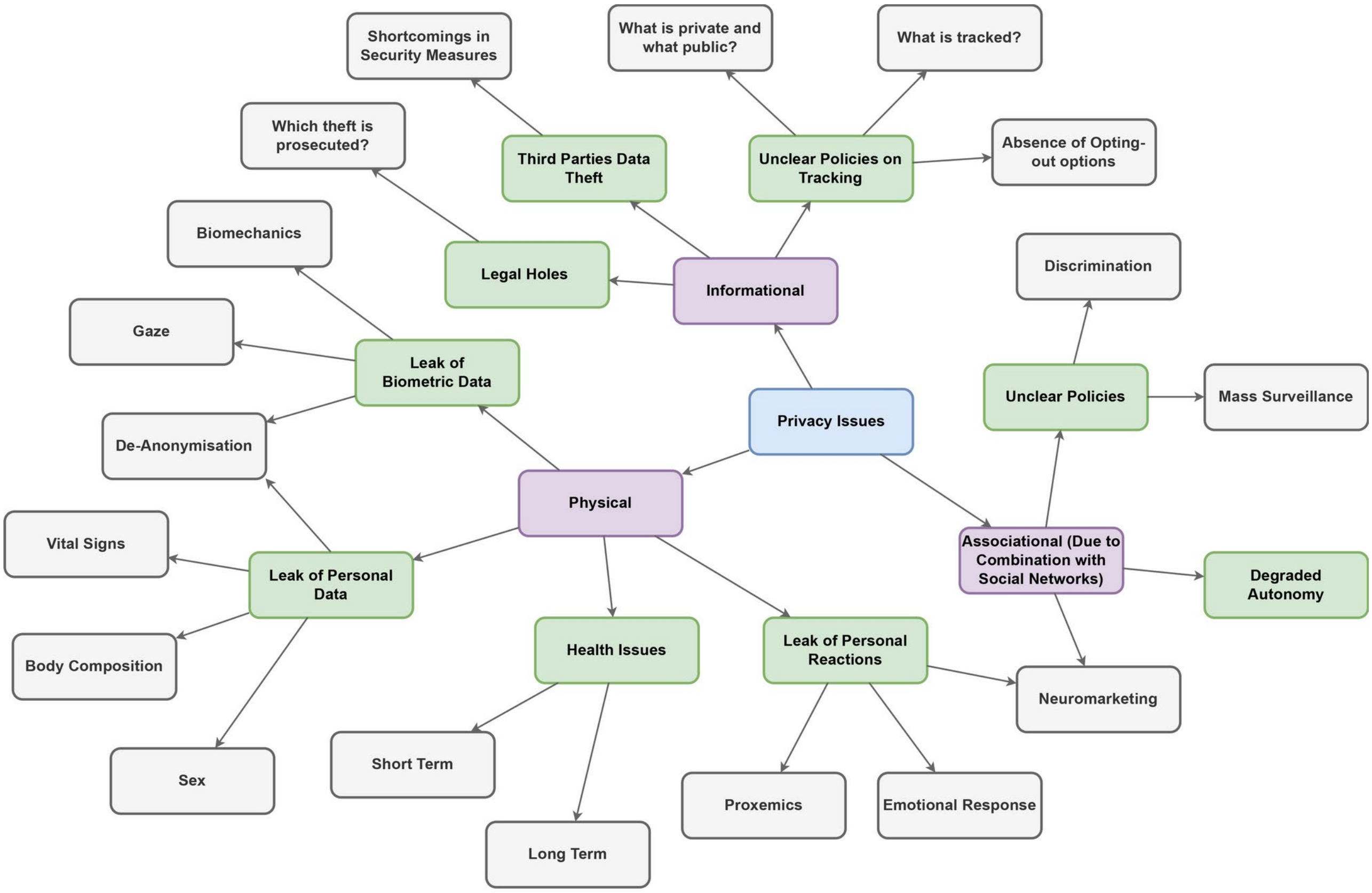

**Fig. 2** Privacy issues in VR. In purple, the three main categories are highlighted. Each category is further categorised into subcategories and specific concerns

that 6 users and 6 developers raised concerns on the data collection practises, while 7 participants out of 20 expressed general mistrust in headset producers. The authors also highlighted that developers show consideration for users' privacy, but that there is a marked disconnection between general concerns and concerns about their own products. One of the interviewees expressed concern for the perception capabilities of the VR sensors and cameras. As shown by Durbin (2017), Oculus Rift sensors are regular webcams, equipped with a firmware and a physical filter that allow infrared light to pass, while removing visible light. However, Kreylos showed that it is possible to operate software manipulations and extract realistic pictures from the Oculus camera, even though it is meant to perceive only the infrared spectrum.

O'Brolcháin et al. (2016) discuss the privacy concerns that derive from the combination of VR and social networks. They divide threats to privacy and privacy issues into three categories:

- Associational privacy, in particular the ability to include and exclude people from certain events or the *global village*, in which everyone could potentially learn important and trivial matters about others;
- Informational privacy, in particular the increased vulnerability of data or its misuse;
- Physical privacy, in particular the prevalence of recording devices or the unintended revelation of physical information (such as physical reactions to ads and the loss of anonymity).

Although this categorisation focuses on the threats arising from the combination of VR and social networks, it applies equally well to general privacy threats to VR. In addition to privacy issues, the authors note that when privacy cannot be ensured, the autonomy of individuals (i.e., their ability to form independent opinions, plans, and goals) is also at stake, because autonomy is intertwined with privacy. Human beings need privacy to think through their problems, make mistakes, and explore different possibilities. Without the guarantee of privacy, users might experience a decrease in autonomy or even lose it.

### 3.1 Threats to associational privacy

In 2016, Nwaneri (2016) noted that Facebook has a track record of experimenting with its users (e.g., showing that removing negative or positive news from feeds affected their mood). They also noted that the Oculus policies were phrased in such a way that it would allow Facebook to conduct similar experiments. The risk to users' privacy would be even higher than the ones found in social networks, given the kind of data gathered by VR systems. In particular, Nwaneri argues that the policy statement regarding collecting "information about [the users'] physical movements and dimensions when [they] use a virtual reality headset" could facilitate discrimination and mass surveillance. According to the author, failing to address these points could lead companies to face serious consequences, such as accusations of violating state privacy and wiretapping laws. On a side note, Nwaneri highlights that the long-term effects of immersive sensory experience granted by VR are still unknown. VR environments might cause safety risks in their users, such as seizures, post-traumatic stress disorder, or interference with childhood mental development. Therefore, it is in the best interest of VR companies to push for clear policies and legislation that regulate their business.

Tromp et al. (2018) elaborated more on the issues of combining social networks and VR, to produce multi-user social VR environments. Among various psychological risks, which fall outside the scope of our work, the authors list privacy. Head-mounted displays (HMDs), due to the sensors they are equipped with, record detailed user's physical and psychological responses to stimuli. These responses could be analysed automatically and used for targeting users with tailored advertising, a method known as *neuromarketing*. In their work, they provided some recommendations to different actors involved with VR applications, such as users, parents, software and hardware companies, regulatory institutions, and researchers. For example, it is still unclear whether VR addiction exhibits similar characteristics to internet use addiction. Therefore, psychologists should establish if the same treatments can be used and VR users should be aware that doctors may not yet know how to treat such addiction.

### 3.2 Threats to informational privacy

Regarding the legislative issues, Lake (2020) highlights that the current US legal framework for online identity misappropriation does not adequately protect VR users. In particular, the author notes that the combination of different laws leaves plaintiffs without any party to sue: anonymity laws make it impossible to sue the identity appropriator and ISP immunity also prevents the victim from suing the VR provider. Until these issues are addressed (for example, following some of the suggestions given by Lake), victims of identity theft in VR environments cannot even bring their claims to US courts.

Kumarapeli (2021) highlights that the major players in VR are now focused on producing realistic avatars, tracking facial expressions and user body movement. On the one hand, this can help to enrich communications in VR. On the other hand, the capability of tracking such information, without allowing users to opt out from undesired tracking, can expose several privacy issues. For example, if the system tracks hints of anger during a VR-mediated collaboration and shows that explicitly on the avatar face, the collaboration could be negatively affected. Therefore, it is important to adopt strategies that allow for enriching avatars' behaviour, while filtering undesired emotions and informing the users about the emotions currently detected.

In 2020, Miller et al. (2020a) wrote that a considerable number of works on user authentication in VR assume a positive connotation. In other words, many papers explore the identification of users as a way for providing seamless and non-invasive authentication. However, the potential shown by VR data to provide authentication implies the potential to identify and de-anonymise users against their will. In their work, they showed that VR motion data can be used to identify users without requiring them to perform any specific action, contrary to what VR authentication studies have done before. The authors recruited 511 participants and showed them five 360-degrees videos (randomly picked from a pool of 80 videos), each 20 s long. The participants did not have to perform any particular task and were free to look around as they preferred. The results of the study show that users can be identified on the basis of their motion data with an accuracy of 87.5%, using a simple random forest (RF) algorithm. The authors also noted that the most important features included height, posture, distance from VR content, focal length, and placement of the hand controllers at rest. In conclusion, just collecting data of VR interactions shows high probabilities for identification, even if such interactions were not designed with identification intentions.

Wierzbowski et al. (2022) took a step further to answer whether it is possible or not to identify users immersed in a 360-degree video. With respect to Miller et al. (2020b), the authors first notice that the previous study do not take into account any body normalisation, meaning that the raw Y-axis value of the headset and the controllers yield lots of information regarding height and arms length of the users. They also noted that VR users tend to stand still in a 360-degree video-watching task, while only their orientation changes. Removing the raw X and Z positional data degrades the accuracy to 19.5%, using Miller et al.'s approach, indicating that their methodology was over-reliant on session-dependent floor position. In addition, Wierzbowski et al. incorporated derivative features (velocity, angular velocity, acceleration, and angular acceleration) that were not used by Miller et al., as well as an eye-tracking sensor. Although they used less information than Miller et al., the authors managed to reach an identification accuracy of 74.7% on a dataset of 43 participants.

Nair et al. (2023) conducted a large study on over 55,000 VR users, showing that it is possible to uniquely identify a user just by leveraging their head and hand motion data. Among the various classification models initially tested by the authors, LightGBM was chosen as the most promising. The experiment carried out shows that, after training on 5 min of motion data for each user, the classification model is capable of identifying a user with 73.20% accuracy when using 10 s of motion as an input, and 94.33% when the model is provided with 100 s of data.

Following the intuition that VR headsets sit tightly on users' heads, Shi et al. (2021) proved feasible an innovative eavesdropping attack, Face-Mic. The proposed attack uses the headset motion sensors, to collect the motion data produced by facial movements, naturally occurring during human speech. Then, the authors developed an innovative deep regression technique to neuter the disturbance of other body movements, allowing them to clean the motion data from non-relevant information. Last, the data is analysed through a deep-learning-based framework, and the results show that, not only it is possible to identify the gender and the identity of the user, but it is also possible to recover parts of the speech content. In particular, these results highlight privacy concerns on VR applications that require users to provide also vocal input.

Given these results, it is apparent that VR sensor data can yield sensitive information and should not be easily available to anyone. Wu et al. (2023) investigated the degree of trustworthiness of VR sensor data, focussing in particular on the SDK APIs of OpenVR, Oculus Platfrom, and WebXR. The experiments, conducted in 2023, show that most onboard sensors required no security permissions whatsoever to access the data, exposing users to data exfiltration. The authors tested a case study in which a victim uses a virtual keyboard in a VR environment to input a password, while an attacker develops a malware to gather the users' movement from their data and infer the password. In their threat model, the attacker has no previous information regarding the victim's VR setup, nor previous labelled sensor data. The experiments show that an attacker can access to sensor data (due to the poor security measures for accessing such data) and infer both passwords and natural language text, with accuracy of 84.9% and 87.1%, respectively. Also (Slocum et al. 2023) analysed the feasibility of striking a side-channel attack to infer key presses on a VR keyboard, this time by analysing the head motion sensors data. The intuition behind the attack is that users' head moves in subtle ways while they type on a keyboard, subtle movement that is enough to infer their inputs. The 21 recruited participants were assigned two tasks: a word typing task and a sentence typing one. The authors propose a classifier named TyPose, split in two modules that are concerned with segmenting sentences and classifying words, modules that leverage CNNs. When training and testing on all participants, depending on the combination of CNNs chosen, the top-1 accuracy results can reach up to 75%.

Lee et al. (2023) investigated indirect keystroke inference attacks on virtual keyboards. In their case, the authors analysed WebVR, an environment for rendering VR applications on regular browsers, and discovered a vulnerability in the APIs. If a VR user has two tabs open in the same browser, one on a WebVR page and the other on a malicious website, the latter can have access to the data of the controllers rendered in the former. As long as both tabs are open, the malicious website can collect all the inputs from the WebVR website, without permission. Using an SVM classifier, the authors show that it is possible to extract sensitive information from such an attack. For example, they infer keystrokes on a virtual keyboard with an average classification accuracy of 93.98% and average identification accuracy of 96.8%.

In light of these considerations, one of the future challenges for VR will be to design non-identifying interactions. For example, manipulating positional data (e.g., altering raw and relative positional data) could obfuscate some physical characteristics and help to protect users' privacy.

Maloney et al. (2020) show that most social VR platforms (such as AltspaceVR, RecRoom, and VRchat) do not clearly inform users about which collected information is private and which public. After conducting a survey among VR users regarding their privacy-related behaviour on social VR platforms, the authors drew some interesting conclusions. First, users showed a similar tendency to share their experiences and emotions in VR, compared to traditional social networks. Second, social VR platforms aim to provide to their users an immersive experience by using resembling avatars. However, this leads to the leakage of some personal information, such as height, race, and appearance in general.

Finally, Zhan et al. (2023) focus on the data minimisation principle, which states that not only vendors should clearly state which data is collected through their applications, but they should also collect (and declare) only the information that is necessary for the application to operate, no more, no less. They verified what is the degree of compliance of VR apps developers through a large-scale study on all 1726 apps offered by Oculus and the authorised third-party App Lab. The results show that 48.1% might be violating the data minimisation principle, by collecting non-critical data, such as email and IP addresses.

### 3.3 Threats to physical privacy

Buck and Bodenheimer (2021) point out that the way that users interact with the surrounding environment can be critical information itself. The authors highlight that research has neglected the relationship between users and personal space in VR environments. In particular, personal space is plastic and varies depending on the interaction at hand: for example, personal space restricts when users approach objects they enjoy and expands when the interaction is not appreciated. This knowledge could potentially be used by companies to gather information about personal preferences, done by positioning an item in the environment and verifying whether users approach it or walk away from it. In turn, this could allow identifying and profiling people without their knowledge nor consent. Moreover, the psychology literature shows that stress increases when interactions happen within personal space, hence personal space should be calculated and taken into account for designing pleasant virtual interactions. In conclusion, personal space is easy to detect and use to infer information about VR users, and this poses the non-trivial problem of how to collect and treat this type of data.

In general, any interaction in VR can lead to private data leakage. Falk et al. (2021) prove feasible a de-anonymisation attack on VR which correlates VR avatars to their human users, leveraging human unique movement patterns. Named ReAvatar, this attack uses only the movement information provided by the VR setup, available without any peculiar privilege or malicious code injection. In their experiments, both the attacker and the victim join a VR game (i.e., VRChat), where people are embodied as avatars and can interact with other people's avatars. The goal of the attacker is to discover who is hiding behind a certain avatar by analysing the movement patterns of the observed victim. To do so, the attacker asks the victim to perform common and apparently innocuous movements, such as dancing or picking up objects. The attacker observes the victim's avatar and extracts the features of their movement. Then, the authors implement pose estimation using part affinity fields (PAF), detect keywise pairs of poses, and extract the local coordinates of *(x,y,z)* from the victim avatar. Finally, each joint pair is evaluated against a threshold that determines if the avatar corresponds to any known person. In a limited study of 6 avatars and 5 users, ReAvatar achieved 89.60% accuracy, proving that the attack is at least feasible.

Gopal et al. (2023) show that the use of VR equipment in public can yield serious risks for users' privacy, demonstrating that it could be possible for a malicious attacker to observe and record a VR user performing tasks in VR and extract sensitive information. In their paper, using a dataset of 368 video clips of VR users, the authors manage to decipher an average of 75% of the users' inputs on a VR virtual keyboard. The extracted information span over different attack scenarios, from graphical pattern lock inputs, to password entry, including even email content entry and text entry on browsers. When the 42 study participants were questioned regarding their perception of VR, the majority believed that VR devices are secure for sensitive transactions and 95% of them were unaware that VR could be vulnerable to similar attacks. Besides confirming the possibility of inferring VR users' keystrokes and hand gestures from their equipment sensor data, Zhang et al. (2023a) show that it is possible to infer also users' voice commands. The paper covers both AR and VR equipment, but for the sake of this paper scope, we discard the former and focus on the side-channel attack on VR. The threat model assumes that a malicious software runs in the background while the VR application runs, allowing it to access to the memory allocation APIs of the VR environment (whether it is Unity or Unreal) and the performance counters (i.e., frame rate, thread data, and rendering tasks. Although each application can, in principle, only access its own memory allocation and performance information, the authors note that the use of shared resources can allow a malicious software to infer data about other applications which require the same resources. By leveraging this knowledge, the authors show that attackers can extract features from time-series data and train classifiers and regressors to identify users' information. Using decision trees (DTs), they achieved a best F1-score of 93.7% when inferring voice commands on Meta Quest.

Carr et al. (2023) discuss how skeleton-based motion capture data, although apparently anonymous, it can still carry some relevant personal identifiable information. They propose and implement the linkage attack neural network (LAN), which works similarly to the Siamese network approach. LAN leverages two semantic-guided encoders that take a target skeleton and a reference skeleton, and then feeds the extracted low-dimensional embeddings to a classifier that tries to assess whether the skeletons match or not. With an F1-score of 74%, the system shows that it is possible to de-anonymise users by using their skeleton data, outperforming other traditional ML algorithms. To counter the issue, the authors propose to use a motion re-targeting approach, in which the system takes the raw skeleton data and cast the motions on a dummy character. By creating this new and standardised dummy skeleton, the authors show that they can provide a high level of privacy, similarly to other anonymisation methods, verified in their paper. However, all anonymisation approaches scored very low in terms of data utility, meaning that anonymisation techniques render almost unusable the skeletons datasets, although the authors' method scored a little higher than the competitors' approaches.

The data provided by VR equipment sensors is rich to the point that it could allow attackers to extrapolate even more sensitive data than what described so far. Zhang et al. (2023b) show that motion sensors embedded in VR equipment can be used to infer users' vital signs. In their paper, they propose an attack named FaceReader, which uses the VR headset motion sensors to detect the subtle facial vibrations produced by users' heartbeat and breathing patterns. These raw vibration signals are used as inputs for a reconstruction network composed of a combination of LSTMs and a self-attention layer. The paper shows that the resulting waveforms, which represent the basic vital signs of the users and are already considered sensitive data, can be used to infer further characteristics, such as users' sex, their identity, and their body fat ratio.

It is also worth noting that eye-tracking technologies are increasingly integrated into modern VR HMDs. Eye-tracking data have the potentiality of revealing more information than it is possible by leveraging headsets motion sensors. For example, Partala et al. (2000) showed that pupil size varies according to emotionally provocative sound stimuli. Their experiments suggested that the size of the pupil discriminates during and after different variations of emotional stimuli. This means that eye-tracking technologies embedded in VR systems might allow extrapolating emotional information, without explicit knowledge or consent from the users. In Sect. 5.3 we cover the most relevant VR authentication methods in the literature that take advantage of eye-tracking sensors.

### 3.4 Solutions for strengthening privacy in VR

Bozkir et al. (2019) proposed a method to assess the cognitive load of users during a driving simulation experience, while preserving their privacy. To do so, they used critical and non-critical time frames, trained multiple classifiers (SVM, DTs, RFs, and KNN with k=1,5,10) and validated their proposal with a leave-one-person-out cross-validation (LOOCV) approach. The authors used a few short time frames, which allows them to predict cognitive load in real time using minimal data, thus preserving users' privacy. Their experimental hardware setup included a HTC-Vive, a Logitech G27 Steering Wheel and Pedals, Phillips headphones, and a Pupil-Labs HTC-Vive binocular add-on for eye tracking.

Regarding the use of eye tracking technology, John et al. (2019, 2020) note that the trend of equipping VR headsets with eye tracking sensors puts users' privacy in danger. Eye-trackers collect images of the users' iris patterns and iris scans, often used as biometrics for a variety of services, can identify people with high accuracy. To counteract this issue, John et al. (2019) showed that Gaussian filters can be applied to iris images to blur iris patterns, making them unrecognisable while preserving gaze detection capabilities. Later in 2020, John et al. (2020) investigated a different approach, hardware-based. Their idea is to attach the eye tracker to a small telescopic arm, allowing users to manually defocus images of the iris at will. In addition to being effective, their approach has the valuable feature of allowing users to directly control when they want to show their iris (e.g., if needed for authentication purposes) or when they want to hide it.

David-John et al. (2022) discuss de-anonymisation attacks against public eye-tracking datasets. On the one hand, public datasets are critical for enabling research on VR regarding users activity and intent; on the other hand, eye-tracking features could enable malicious users to analyse public datasets and re-identify study participants. To mitigate this risk, first the authors adapted the definitions of k-anonymity and plausible deniability for the specific eye-tracking application and applied it on 7 public-available datasets. Then, they evaluated the performance of a hypothetical de-anonymisation attack through a radial basis function (RBF) classifier, showing that the chances of re-identification go from almost 100% to 30% with exponential-DP, which is the random chance rate. At the same time, the accuracy is preserved to 72%, showing that this approach does not render the dataset unusable. The authors make an important remark, for what it concerns the need of striking a reasonable balance between privacy and usability requirements.

A similar remark can be found in the work of Wei et al. (2023), which focus on the collection of a dataset of 3D virtual worlds with 6 degree-of-freedom, for enabling other researchers' investigation on utility-privacy tradeoffs in VR privacy-preserving technologies. The dataset is composed of 2 outdoor and 2 indoor scenarios, collected by 31 different participants that were free to roam and interact with virtual objects. The collected features include headset position and orientation, controllers position and orientation, controller keystrokes, eye gaze, virtual objects position and orientation, RGBD of the physical participants during the experiment, and follow-up questionnaires.

Ren et al. (2024) proposed DPGazeSynth for facing the same issue: strengthening privacy protection of eye-tracking data without degrading its utility. The approach of the paper allows synthesising gaze paths and distinguish between fixations and saccades, and uses a Markov chain model to preserve correlations and enhance analysis accuracy. The resulting model provides formal differential privacy guarantees, concerning data collection and analysis. Although some other solutions have been proposed in the literature for eye-tracking data, such as the state-of-the-art (Li et al. 2021), those solutions exhibit limitations in VR scenarios, unlike DPGazeSynth.

Also Wilson et al. (2024) investigate the challenge of mitigating de-anonymisation (or re-identification) attacks on eye-tracking datasets in VR, while ensuring data utility at the same time. However, while other works discuss the data-centric utility, which is the preserved degree of utility that a dataset has in the interest of downstream processes such as users' gaze prediction, Wilson et al. focus on the user-centric utility. In other words, they investigate methods that allow for granting users' privacy without negatively impacting their VR interactiveness experience, pointing out that VR users would be less likely to opt-in for a privacy-preserving mechanism that would render their VR experience less enjoyable (e.g., by introducing stuttering). The authors evaluate three different privacy mechanisms previously proposed by David-John et al. (2021) for traditional eye-tracking use cases, (i.e., additive Gaussian noise, temporal downsampling, and spatial downsampling) and a novel privacy protection mechanism, the smoothing operation. First, it is worth mentioning the manuscript shows that, while the baseline risk of de-anonymisation in VR eye-tracing is lower than with conventional eye-tracking, there is still a considerable risk for the users (up to 67.31% of re-identification accuracy). Regarding the four possible countermeasures proposed, the results on a pool of 8 participants show that spatial downsampling and smoothing operation can be viable privacy-preserving mechanisms. Most importantly, both approaches retain high subjective usability, hence exhibit minimal impact on users' experience.

In 2022, Trimananda et al. (2022) studied Oculus VR platform and its apps ecosystem, focusing on the consistency between collected data and stated privacy policies. The paper describes OVRSeen, a methodology for collecting, analysing, and comparing actual network traffic emerging from VR apps, and the official privacy policies. For any given Oculus app, OVRSeen allows for capturing and extracting network data, reconstructing the actual data flows, and comparing them against the privacy policy statements on the app store. On a dataset of 140 apps, the authors found that 70% of the apps exhibited dataflows that were either not disclosed or consistent with the privacy policies. Of these 140 apps, 38 apps did not even state privacy policies, and many app developers neglected to declare data collection by the platform and third-parties. Interestingly, when the authors attempted to include the privacy policies of third-party apps into the main analysed app, then 74% of the dataflows became consistent, showing that many issues with network-to-policy consistency stems from a lack of transparency in which data is shared with third-parties and how third-parties use the data. Last, OVRSeen has also highlighted how 69% of the dataflow corpus is unrelated to the apps core functionalities, but serves other goals, such as advertising, marketing, and analytics.

More recently, Guo et al. (2024) developed VR-SP, a privacy and security assessment tool for VR applications which integrates static analysis techniques and privacy-policy analysis methods. When testing the tool against 500 popular apps offered in the Meta Oculus and Quest 2 app stores, the authors detected a number of privacy leaks and security vulnerabilities, including the use of insecure hash functions and encryption algorithms, the lack of compliance with users' permissions to collect their data, and the lack of data flows checks. Besides, the VR-SP tool highlighted various inconsistencies between the privacy policies statements of the apps and the actual data collection, similarly to what has been highlighted by Zhan et al. (2023) as previously discussed.

Sakuma et al. (2024) discussed a relatively unconventional use of VR for allowing people to cooperate online by reproducing in VR the real environment surrounding the users. Through the use of point cloud LiDAR cameras, the authors foresee the capability of live reproducing in VR the environment and the effects of the physical interactions within such environment. However, they also point out that this could create new privacy-related issues, such as the reproduction of objects or parts of the environment that the user might wish to exclude from the live rendering. Therefore, they propose the use of a privacy filtering system that allows individuals to exclude elements at will and, ideally, learns to automatically exclude elements based on previous preferences.

Related to what has been discussed by Gopal et al. (2023) in the previous section, Yang et al. (2024) study inference attacks on VR keyboard typing. The paper investigates inference attacks struck both by direct observation of VR sensors' data and by observation of the avatar's finger movements within the virtual environment. After showing, unsurprisingly, that it is indeed possible to strike such attacks, the paper provides two suggested mitigations. First, the authors suggest that the VR system should be programmed to monitor users' actions and, once detected the act of typing on a keyboard, it should severely limit access to the hand-tracking API by any application. For example, by banning access, or imposing a significant limit on query frequency, or largely reducing the sampling rate of the hand-tracking sensor. The second suggestion is to add noise to the sensor data, in particular to perturb the $y$ depth data. In the paper, the authors experiment with a moderate and high zero-mean Gaussian noise added to the $y$ value, showing that moderate noise is already effective at increasing the word error rate (WER) from 14.2% to 55.4%. High noise levels increase the WER up to 71.4%.

Last, concerning the use of head motion data for exfiltrate information regarding VR users' typing on virtual keyboards proposed by Slocum et al. (2023), the authors also discuss possible mitigation strategies. After showing that reducing the sampling rate of the headset motion sensors (i.e., accelerometer and gyroscope) and reducing the data precision by rounding off decimals are not effective strategies, the authors suggest possible future directions. First, they suggest that the keyboard typing task is always transferred to a separate virtual room, such that malicious apps would not have a chance to access the sensors data—all of this, at the expense of users' immersion. As an alternative, the position, size, and keycaps position (in case of PIN-insertion tasks) could be randomised. However, none of these mitigation strategies have been tested by the authors and more research is needed.

## 4 Taxonomy of threats to VR security

Adams et al. (2018) interviewed 20 participants, 10 users and 10 developers, investigating their perception of security in VR. Only 2 users raised concerns about malicious applications, 4 participants in total worried about data sniffing, and 3 users said that security ”is not there yet”, which suggests that not many end-users might have cybersecurity concerns. Given this attitude, the authors are of the opinion that many early VR applications will exhibit different security vulnerabilities, as cyber-security does not hold a high perceived value for end-users.

As previously mentioned, a recent study from Guo et al. (2024) showed how the conclusions from the earlier study of Adams et al. (2018) (conducted 6 years prior) are incredibly actual, with dozens of insecure apps currently offered in popular VR app-stores. Yeboah-Ofori and Hawsh (2023) call attention to the fact that VR has the potential of being a critical technology for disabled users and their accessibility, education, rehabilitation support, and social inclusion. Security vulnerabilities that undermine proper functionalities can undermine the safety and well-being of disadvantaged users, more than it could impact non-disabled users. Jones et al. (2023) highlight the fact that, due to their traits, neurodiverse users (e.g., users affected by ADHD, autism, or dyslexia) could be particularly sensitive to some security and privacy threats, such as immersive attacks and perceptual manipulations.

Therefore, it is critical to assess which attacks could be expected and mounted on VR. In this section, we provide an overview of the different attacks that can affect VR, categorised according to the tree shown in Fig. 3. First, we discuss VR-specific attacks that are made possible by the very nature of VR and the involved sensors. Then, we cover

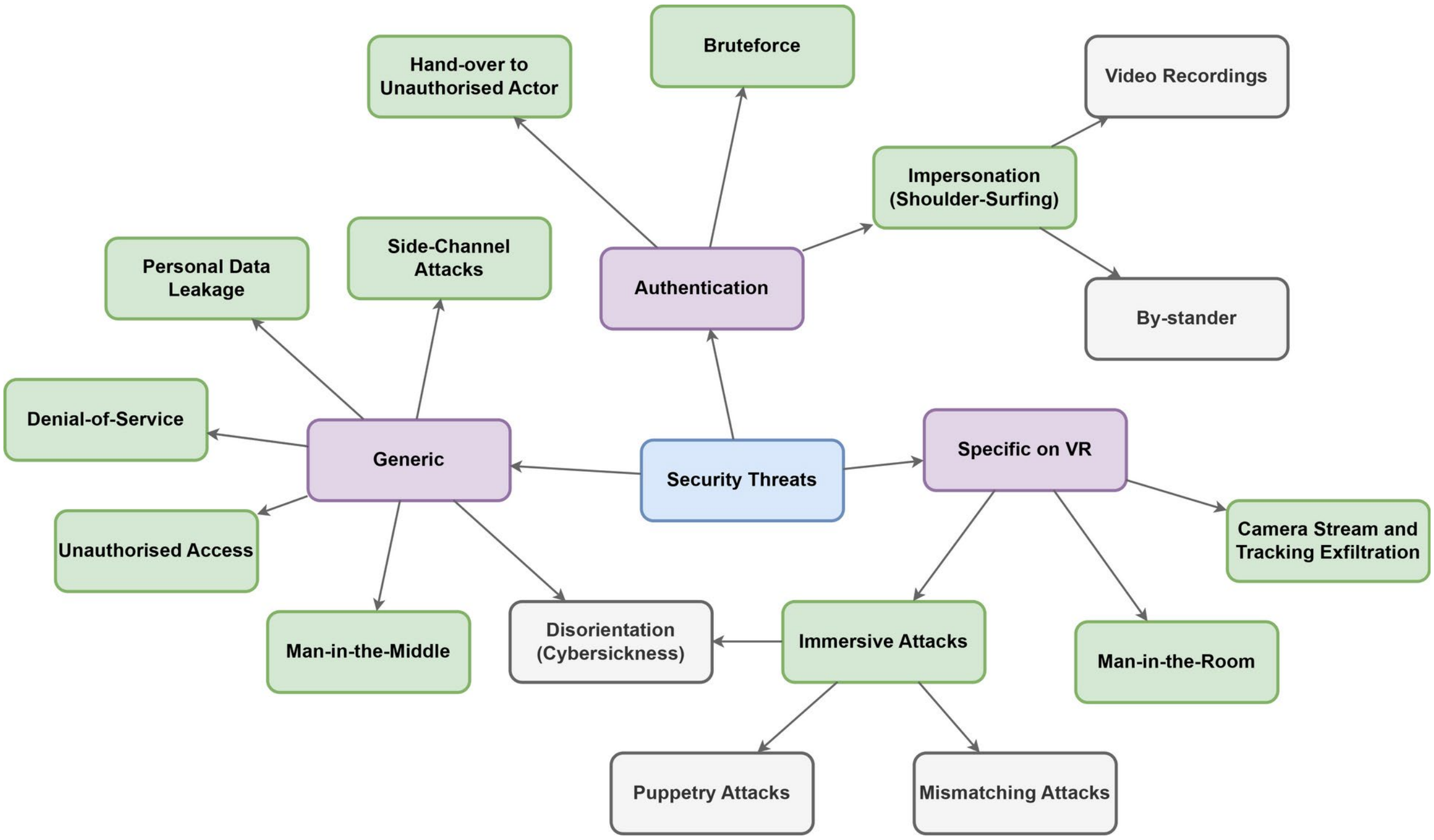

**Fig. 3** Security issues in VR. In purple, the three main categories, further specified in subcategories and specific concerns

the generic security issues that can affect VR, as they affect various information systems.

## 4.1 VR-specific security threats

Tseng et al. (2022) note that, due to the immersive nature of VR, malicious attackers could exploit virtual-physical perceptual manipulations (VPPMs) to harm VR users. VPPMs, a set of techniques that allow to expand users' interaction with a VR environment, include various techniques. Examples include the modification of users' arms movement to provide haptic feedback, or the guided redirection of users to a larger physical area to improve their experience. Although useful, these techniques could be exploited by malicious attackers to inflict physical harm on targeted users. The authors classify these attacks in two main categories: puppetry attacks and mismatching attacks. By their definition, puppetry attacks "control physical actions of different body parts of an immersed user". Mismatching attacks are "manipulations in which the adversary exploits a difference of information between a virtual object and its physical counterpart to elicit misinterpretation for the VR user". In their paper, the authors show how different VPPMs can be easily subverted for creating physical harm to VR users and bystanders. For example, by tilting the virtual environment and make a person walk towards the stairs instead of a door, making them fall. Or by drawing a chair that does not exist in the room and tricking the user into sitting, hence falling and getting hurt. Depending on the specific attack included in these categories, both puppetry and mismatch attacks could have different impacts in terms of the Confidentiality, Integrity, Availability, and Safety (CIAS) quartet, as shown in Table 1.

Also (Casey et al. 2021) noted that VR, due to its immersive nature, can have a considerable impact on people. Therefore, security is of paramount importance. In their paper, the authors formulate four VR-specific attacks, classified as Immersive attacks, as they leverage the unique immersive characteristics delivered by VR systems. These attacks are:

– Chaperone attacks manipulate the Virtual Environment (VE) boundaries (i.e., tampers with the walls drawn in the VR scene) to make the collision avoidance fail and put users' safety at risk;
– Disorientation attacks aim to cause dizziness and confusion in VR users, causing a condition referred to as cybersickness. Dennison et al. (2016) define cybersickness as a form of motion and simulation sickness due to physiological factors of the users, in correlation with their immersion and presence in VR environments;
– Human Joystick attacks aim to control the physical movement of a user, such that they move to a predefined spot without realising it;
– Overlay attack, where an attacker inserts in the VE unwanted objects, such as images and videos.

**Table 1** VR-specific attacks and their impact in terms of the CIAS Quartet

| | Confidentiality | Integrity | Availability | Safety |
|---|---|---|---|---|
| Mismatching attacks | ✗ | ✗ | ✗ | ✗ |
| - Chaperone attack | | ✗ | | ✗ |
| - Overlay attack | | ✗ | ✗ | |
| - Man-in-the-room attack | ✗ | ✗ | | ✗ |
| Puppetry attacks | | ✗ | ✗ | ✗ |
| - Human joystick attack | | ✗ | ✗ | ✗ |
| Disorientation attack | | ✗ | | ✗ |
| Camera stream and tracking exfiltration attack | ✗ | | | |
| Side-channel attack on virtual interactions | ✗ | | | |

Different attacks can impact different component of the quartet, therefore the two categories proposed by Tseng et al. (2022) include all the components that could be separately impacted

In the context of the two main categories provided by Tseng et al. (2022), we can categorise the human joystick attack in the puppetry attacks. The chaperone attack and the overlay attack, instead, can be classified as mismatching attacks.

In addition to the four Immersive attacks, Casey et al. (2021) list two additional possible attacks. The first is the Camera Stream and Tracking Exfiltration attack, where the attacker gains access to the HMD live stream and the HMD front-facing camera stream. The second is the Man-In-The-Room attack, which consists of an attacker joining the target's VR environment while remaining invisible, allowing to extract information regarding the user or maliciously manipulate the environment. Regarding the two categories previously proposed by Tseng et al. (2022), this attack can be categorised as a mismatching attack and could lead to confidentiality threats. In general, these attacks have different impacts in terms of the Confidentiality, Integrity, Availability, and Safety (CIAS) quartet, as shown in Table 1.

The authors analysed OpenVR, an SDK developed by Valve, to identify how an attacker could perform the attacks that they identified. Using an HTC Vive headset and an Oculus Rift, the authors proved that these attacks could be carried out. They found that Chaperone, Disorientation, and Human Joystick attacks could be struck by modifying a JSON configuration file. For the Chaperone attacks, the authors modified the virtual room setup stored in the JSON file. To strike a Disorientation attack, they modified two parameters that control players' orientation, stored in the same JSON file. The same parameters were manipulated to carry out a Human Joystick attack, even though the sequence and the rate of modifications are different. They also proved possible the Overlay attack, due to the fact that any OpenVR application is allowed to call an overlay and that no mechanism to force closing overlays existed. The authors suggested that this could be exploited to show undesired advertisements or even perform ransomware attacks that prevent users from using their VR system. Last, the authors were able to extract information from the video buffer of the HTC Vive; the Oculus Rift was not used as it was not equipped with a front camera. In conclusion, every attack was successful, and the reason was OpenVR lack of robustness: critical information was stored in plain-text, permissions were not defined or not restrictive enough, and applications were insufficiently isolated without sufficient integrity checks. Casey et al. suggest that Arya (Lebeck et al. 2017, 2018), a policy enforcer created for AR applications, could be adapted to the VR domain.

Regarding Overlay attacks, Ahn et al. (2018) noted that most of the research conducted for AR and VR security has focused more on input privacy (i.e., minimising the sensor data available to third parties) than visual output security. A notable exception is the aforementioned Arya (Lebeck et al. 2017, 2018), proposed by Lebeck et al. The authors showcase that defining rule-based policies for every conceivable AR/VR scenario is an intractable task, using a notable example that assume an application required to create various virtual objects, which should not obstruct the users' view nor overlap with other important objects (e.g., a virtual avatar). In other words, in this application, a new dynamic object should be positioned in the environment in such a way that it is visible but not obstructive. To overcome the complexity of managing this task, the authors proposed Arya, which uses Fog computing and Reinforcement Learning to automatically synthesizes security policies. As suggested by Casey et al. (2021), their approach can be used to prevent attackers from manipulating a VR application, either to distract and disrupt users or to prevent expected interactions by obstructing important environmental objects. An example of an attack would be a ransomware that obstructs the user's view with an object until the victim pays a ransom.

Last, VR platform might be vulnerable to various specific side-channel attacks. For example, Al Arafat et al. (2021) investigated keystrokes on virtual keyboards in terms of side-channel security threats. The authors proposed VR-Spy, a side-channel attack that allows to extract gesture patterns from channel state information (CSI) wave-forms of Wi-Fi signals and then, by applying machine learning, recognise keystrokes from such patterns. The experimental setup consists of a person equipped with a commercially available headset, positioned between 2 commercially off-the-shelf (COTS) wireless devices. Their experiments show that VR-Spy achieves an accuracy of 69.75% when performing virtual keystroke recognition. Referring to the CIAS quartet, this specific side-channel attack (listed in Table 1) breaches communication confidentiality.

### 4.2 Generic security threats that affect VR

Beyond security threats that affect specifically VR due to its characteristics, VR systems can also be vulnerable to generic security threats. Valluripally et al. (2021a, 2020, 2021b) focused on modelling and controlling generic attacks that could cause cybersickness in VR users. In their first paper (Valluripally et al. 2021a), they introduce a rule-based framework that controls cybersickness by means of 3Qs adaptation. In other words, they propose a system that regulates three aspects:

- Quality of Application (QoA), a measure of application performance;
- Quality of Service (QoS), a measure of network resources such as bandwidth and jitter;
- Quality of Experience (QoE), a measure of user perceived satisfaction.

The framework is useful for reducing sickness, both if induced by incidental performance issues (e.g., temporary network bottlenecks), and if caused by malicious attacks, such as DoS. Through continuous monitoring of anomaly events, the framework classifies events into predefined categories and selects the appropriate adaptation strategy. For example, assuming that the system is deployed on Amazon Web Services (AWS), if the framework detects an abnormal high CPU utilisation that affects QoA, an adaptation strategy would be to upgrade the server instance to a better instance (e.g., from AWS t2.micro to AWS c4.large). Their experiments show that adaptations for QoA anomalies can reduce cybersickness by up to 26.43%, while, in the case of QoS anomalies, the proposed adaptation reduces cybersickness by 30.28%. The framework also has the ability to take into account the associated hourly costs for each adaptation strategy, so that the best cost-effective adaptation can be adopted.

In their subsequent work, Valluripally et al. (2020, 2021b) expand on their analysis of sickness-inducing attacks. First, the authors conducted an experiment where they simulate the effects of ongoing security attacks and/or privacy attacks while users perform some tasks in VR. The experiment was performed by artificially manipulating the traffic rate, and it shows that these attacks cause considerable disruption to different factors that contribute to cybersickness, such as nausea and discomfort. The authors then proved that cybersickness can be correlated with security and privacy issues, regardless if due to attacks, faults, or a combination of both. In particular, they proposed a framework based on an attack-fault tree (AFT) formalism for producing threat models that describe the possible security and privacy attacks. The resulting AFTs are used to perform a quantitative assessment of the probability of disruption caused by each threat. Evidence shows that the threats that induce higher levels of cybersickness in VR-based Learning Environments (VRLEs) are DoS attacks, data leakage, man-in-the-room attacks, and generic unauthorised accesses. Based on the quantitative results, the authors also experimented the effect on cybersickness when different design principles are applied, such as hardening and redundancy. For example, applying hardening and the principle of least privilege reduces the probability of cybersickness by 28.96%, whereas applying redundancy and the principle of least privilege achieves only a small reduction of 3.05%. When all prescribed principles are applied together, the probability of experiencing cybersickness is reduced by 35.18% in total.

Gulhane et al. (2019) focused on a sub-area of VR, namely, VRLEs. VRLEs are VR spaces designed for delivering virtual learning experiences to their users. For doing so, VRLEs often combine data from emotion tracking sensors with other sensitive information, such as users' learning progresses and personal data. Therefore, VRLEs pose a number of challenges with respect to the security, privacy, and safety of their users. Gulhane et al. proposed a risk assessment framework that uses attack trees to formalise internal and external vulnerabilities for VRLEs, such as DoS, SQL injections, and unauthorised logins. For example, they use attack trees to formalise the observation that delayed packets (e.g., due to DoS attacks) would lead to compromise the availability of the server, hence its security. The output is a risk score for each analysed property: security, privacy, and safety. Using vSocial (Nuguri et al. 2021) (a cloud-based VRLE designed for young people with learning disabilities) as a testbed, the authors evaluated the attack trees produced by their methodology. Each leaf of every attack tree represents a threat and, for each leaf, the authors simulated the corresponding attack. Then, they assigned an impact score for every threat and used the scores for computing complete attack trees.

Valluripally et al. (2023) further investigate VRLEs security and privacy issues. First, the authors focused on two categories of general attacks that could disrupt the users' immersiveness experience (i.e., unauthorised access and DoS). Then, they modelled these attacks by using attack trees and proposed an anomaly-detection approach based on the combination of KNN and Z-score-based analysis. Using vSocial as a case-study environment, they recruited 15 participants and showed that the attacks are indeed disruptive to users' experience, as theorised. In addition, they also showed that the proposed anomaly-detection is capable of detecting an ongoing attack within 3.2 s from starting time and a classification accuracy of 97.5%. The authors also tested the performance in case of multiple ongoing attacks, showing that the detection happens within 2.8 s and that the classification accuracy stabilises around 87.5%.

Finally, it should be noted that VR equipment could facilitate security threats, such as shoulder-surfing attacks where a malicious party spies on the user's actions to steal their password. Chen et al. (2018) described a computer vision-based attack, struck using VR augmented with additional sensors. In their experiments, the authors showed that it is possible to use an HMD, equipped with an additional ZED stereo camera, for recording the surroundings and stealing the passwords input by users on their devices. The paper shows that it is possible to reconstruct a 3D video from the two side-by-side videos recorded by the ZED camera. Then, the 3D video can be analysed and used for extracting the 3D trajectories of the fingertip movements to recover the passwords tapped by the victim. The paper shows that there is the potential risk of malicious users that pretend playing some video games on VR while recording unconscious people moving in the surrounding environment. In particular, the experiments show that there is a strong correlation between the success rate of the attack and the distance from the victim, as well as with the number of input attempts. In addition, some authentication systems require confirming the input password, and such requirement increases the success rate of the proposed attack.

**Table 2** Works on authentication for VR

| Authentication approach | Works in the Literature |
|---|---|
| Traditional and VR-interaction-based authentication | George et al. (2017), Yu et al. (2016), Olade et al. (2020b), Funk et al. (2019), Khamis et al. (2018), Mathis et al. (2020b, 2021b), Funk et al. (2019), Rupp et al. (2024), Abdelrahman et al. (2022), Shi et al. (2023), Yang et al. (2023), Turkmen et al. (2023), Huang et al. (2023), Han (2023) |
| Biomechanics | Mathis et al. (2020b, 2021b), Funk et al. (2019), Rogers et al. (2015), Sivasamy et al. (2020), Mustafa et al. (2018), LaRubbio et al. (2022), Riyadh et al. (2024), Pfeuffer et al. (2019), Kupin et al. (2019), Ajit et al. (2019), Liebers et al. (2021a), Miller et al. (2020b), Miller et al. (2021), Olade et al. (2020a), Shen et al. (2019), Li et al. (2024a), Miller et al. (2022b), Wang et al. (2023), Ritola et al. (2023), Deng et al. (2023), Suzuki et al. (2023) Miller et al. (2022a) |
| Eye-tracking | Olade et al. (2020b), Mathis et al. (2020b, 2021b), Liebers et al. (2021b), Luo et al. (2020), Khamis et al. (2018), Lohr et al. (2018, 2020), Zhu et al. (2020), Tricomi et al. (2023), LaRubbio et al. (2022), Lohr et al. (2023), Ritola et al. (2023) |
| Other works | Miller et al. (2022c), Mathis et al. (2020a), Wang and Gao (2021), von Willich et al. (2019), Progonov et al. (2023), Cha et al. (2022), Li et al. (2023), Grandi et al. (2023) |

Concerning the eye-tracking papers, some of those might have also included biomechanics information, such as head position and orientation

## 5 Authentication in VR

Among the areas of research within VR security, the topic of authentication deserves a section of its own. In fact, most of the work related to VR security focuses on identifying or authenticating its users: as the terms are often times used interchangeably, we also make no distinction for the scope of this paper. In this section, we provide an in-depth exploration of the literature, and we summarise the works in Table 2. It is worth noting that, for what it concerns the eye-tracking category, some papers only use eye-tracking data while others combine it with biomechanics features. Since the literature shows that the eye-tracking data is the most important future (by a relevant margin), we deem appropriate to list these papers in the eye-tracking category.

### 5.1 Traditional and VR interactions-based authentication

In their seminal work, George et al. (2017) proposed a first step toward integrating traditional authentication approaches into VR, in particular PINs and swipe patterns. They conducted a study with a total of 30 people and showed that both techniques are suitable for transposition into VR. In particular, they showed that the usability in VR matches the usability in the physical world. Furthermore, they showed that the private visual channel granted by VR makes such authentication processes harder to observe and steal.

Similarly, Yu et al. (2016) investigated the implementation of two traditional authentication methods, pattern swiping and PINs, while also making a first attempt to create a 3D password for VR. Although the authors did not introduce sufficient details about the 3D password, they provided some intuition about the relative resilience of the three methods against shoulder-surfing attacks. In their experiments, the participants were recorded while using the three authentication methods and then asked to guess the passwords of other participants, watching the recordings. The results showed that 3D passwords were considerably harder to guess than 2D swipes and PINs, most probably due to the additional complexity added by introducing a third dimension.

Olade et al. (2020b) further investigated the portability of swipe patterns from mobile devices to VR, noting that George et al. (2017) did not provide a direct performance comparison between mobile devices and VR. In their experimental setup, the authors implemented the swiping authentication system in terms of 4 different interaction methods: using the hand-held controller (HHC), the HMD, the LeapMotion (a hand-tracking device), and the aGlass (for eye tracking). While the mobile swiping authentication was the easiest and fastest to use, the HHC and the LeapMotion versions were comparable both in terms of speed and usability. Regarding shoulder-surfing attacks, given the same amount of information, attackers have a considerable lower success rate when swiping is done in VR than when it is done on a mobile phone screen.

Mathis et al. (2020b, 2021b) proposed RubikAuth, a three-dimensional authentication method, consisting of a five-faced cube (one face is omitted for improved reachability) that exhibits 1 colour and 9 digits per face. For authenticating themselves, the user rotates the cube with the left HHC and selects n digits as their chosen password. The entropy of a RubikAuth password is equal to $45^n$, considerably higher than the $10^n$ entropy granted by traditional numerical pins. The authors evaluated the impact on performance of different input techniques, concluding that controller tapping is faster than head pose and eye gaze, while the study participants did not express any clear preference regarding the usability of each input approach.

The authors also evaluated the security of their proposal, using three realistic attackers with different tools: pen and paper to annotate observations, a 3D replica of the cube, and recordings of users performing authentication. The experiments show that having access to such resources slightly increases the chances of guessing a RubikAuth password (and, reasonably, the same applies to other VR authentication modes): 8 attacks out of 540 attempts succeeded, even though the passwords were simple patterns involving only a single cube face. RubikAuth exhibits strong guarantees against shoulder-surfing attacks, as it is possible to apply fake face switches to deceive bystanders. It is also possible to strategically angle the cube and pick numbers from different faces without applying any rotation.

Funk et al. (2019) proposed LookUnlock, a head-gaze-based solution to enter passwords on HMDs. LookUnlock can use different kind of spatially-aware passwords, namely, spatial passwords, virtual passwords, and hybrid passwords. To set a spatial password, users move the HMD cursor to different positions in the environment and perform enter actions. For the unlocking phase, users insert their password by placing the head-gaze cursor over the spatial targets previously defined at the setting phase. After the first object, the users have up to 3000 ms for detecting the next object, before getting a time-out. LookUnlock provides an additional layer of security to the authentication phase by binding spatial passwords to the environment. In practical terms, the same actions performed in a different environment would not authenticate the user, as the framework would identify that the physical environment differs from the one used at enrolment time.

Rupp et al. (2024) created a novel gesture-based authentication system that leverages the interaction with virtual avatars. The VR user is immersed in a virtual office room, populated by virtual agents that act as interaction partners. The agents activate upon proximity of the user in the virtual space, and acknowledge the user presence by establishing eye contact. Once contact has been established, the users can authenticate themselves by performing a set of predefined natural gestures (such as waving, bumping fists, finger-guns, etc.), where each agent requires a different set of gestures. The short paper employed only five participants, but the preliminary results show that their agents-mediated gesture-based authentication could be a viable approach to authentication with shoulder-surfing attack prevention.

To counter the risks of observation attacks struck by bystanders, Abdelrahman et al. (2022) proposed CueVR, a cue-based authentication scheme for VR. The system is based on the PIN paradigm known to every modern IT user, where the numbers on the PIN pad are split in two groups and each group is assigned a colour. In addition, every number is randomly assigned a cue (i.e., a different motion that the user has to perform to select such number). In order to insert their pre-selected PIN, the user must first select the colour of the number to select, and then they have to perform the motion cue associated to that number (whether via controller movement or by operating the integrated trackpad). This randomicity makes it harder for a bystander to guess the PIN, as observers do not have direct access to the colours and the cues displayed in the VR environment. Based on the experiments performed with 20 participants, CueVR is as fast as existing authentication schemes, reasonably useable, and resilient to observational attacks.

#### 5.1.1 Proposed but unvalidated authentication approaches

The following short papers have proposed interesting ideas for novel authentication approaches. However, they lack experimental support and the judgement on their effectiveness is suspended. With NELI-AUTH, Shi et al. (2023) propose another virtual interaction-based approach for authentication, where the user has to move the controllers and rotate the sticks to replicate the appearance of a sequence of white-and-grey spheres. Due to the fact that these spheres (items) are composed of different number of elements, different sequences can be achieved with very similar motions, making it difficult for bystanders to infer the password. Although interesting, the proposal has not been verified through any preliminary experiments and resilience against shoulder-surfing is not proven.

Another proposal that has not been field-tested with practical experiments, is Pathword by Yang et al. (2023). In this paper, the VR user is presented with a replica of planet Earth, on which 50 countries and cities are highlighted. The user can pick a password by choosing freely a sequence of countries and cities. The globe can be rotated with the left hand, while the right hand acts as a pointer that allows to select the cities and the countries. By rotating the globe and selecting the different points, the user composes the password and authenticates on the system. Given that the user chooses independently the authentication sequence and is allowed to rotate the virtual globe in any direction, the authors theorise that their approach should be resilient to shoulder-surfing attacks. Turkmen et al. (2023) proposed a voxel-based authentication scheme, where users upload 2D images of their choice, which are automatically converted into 3D objects (i.e., voxels) in the VR environment. Once the images are converted, the user chooses 4 voxels to create their password. The voxels are randomly positioned in the VR environment and the user will have to select them in the correct order to authenticate themselves, through eye-gaze and controller-pointing.

Detection-Based Authentication (DBA) by Huang et al. (2023), proposes to show users a sequence of 4 different environments and let them pick a different direction either via physical turning or by using the controllers. The different directions per each location are the secret knowledge used in this scenario; preliminary tests ran by the two co-authors, show an average entry time around 8 or 9 s per authentication event. Han (2023) proposes NinjaLock, a system that uses the hand-recognition features of Meta Quest 2. First, for enrolling a password, the users bind 6 different 3D symbols to one hand, either left or right, and a specific hand gesture (23 are listed, as they are easily recognisable and distinguishable by the headset). At login moment, the users are presented in the virtual environment with 4 out of the 6 symbols, and they have to perform the correct gesture with the required hand. To add a confounding variable and diminish the chances that a bystander would learn the different gestures, the user can perform a random gesture with the other hand at the same time.

### 5.2 Biomechanics-based authentication

As pointed out by Miller et al. (2021), traditional authentication methods, such as PINs and passwords, not only can be stolen by malicious attackers, but they also allow a legitimate user to hand-over control to a third party. This can be particularly problematic for VR applications that must ensure that the user performing the tasks is the same user who authenticated themselves in the first place. For example, a VR scene used to undertake university exams should prevent a student from cheating by logging in and handing over the control to a fellow student. One solution for countering handover threats is to implement implicit and continuous authentication via biometrics, which allow for detecting automatically when the user's characteristics differ from the expected ones.

Another downside of explicit authentication, as observed by Liebers et al. (2021b), is that it negatively affects user immersion in VR. Biometrics-based authentication techniques also have an advantage over explicit authentication methods in this regard, as they are implicit and provide a smoother experience for the users. As shown by Riyadh et al. (2024), users with low VR familiarity favour gesture-based approaches over traditional approaches, rating the former method considerably more usable than the second. Besides usability scores, their 40-participants experiment also shows that gesture-based login times are considerably faster for inexperienced VR users, and marginally faster even for experienced users.

In this section, we describe the state of the art for what concerns the authentication of VR users using biomechanical biometrics. Jain et al. (2004) noted that body movement can be used for authentication purposes when it is universal, distinctive, repeatable, and collectible. Based on such observations, several works in the literature explored how VR sensors can be used to capture movement patterns. In this section, we explore the state of the art; in Table 3 we summarise the features used by each analysed work and the accuracy results obtained by the authors.

Li et al. (2016) noted that adults exposed to fast beat audio tracks exhibit unique patterns of head movement. Their observation resulted in Headbanger, a framework that captures unique head movements that arise as a natural response to fast-tempo audio stimuli, and authenticates the users accordingly. Implemented using a Google Glass device, Headbanger shows an EER of 4.43% when users are exposed to a 10-second music cue. When tested against imitation attacks, less than 3% of attempted attacks achieved a false positive.

With VRCAuth, Sivasamy et al. (2020) also investigated the use of head movements to provide continuous authentication in VR settings. VRCAuth performs a binary classification to determine whether the current user is authorised or not. The authors evaluated five different binary classifiers on two different data sets, the *Kaggle VR driving simulator* (Jafarnejad 2017) and the *user behaviours in spherical*

**Table 3** Biomechanics-based authentication in VR without the use of eye-tracking sensors

| | Participants | Accuracy (%) | Equal error rate (%) | F1 (%) | Head position | Head orientation | Controllers position | Controllers orientation | Distance between devices | Body normalisation | Velocity/Acceleration | Gait estimation | Head reverberation | Hand data |
|---|---|---|---|---|---|---|---|---|---|---|---|---|---|---|
| Li et al. (2016) | 30 | | 4.43 | | ✓ | ✓ | | | | | | | | |
| Sivasamy et al. (2020) | 40/48 | 99 | | | ✓ | ✓ | | | | | | | | |
| Mustafa et al. (2018) | 23 | | 7 | | ✓ | ✓ | | | | | ✓ | | | |
| Pfeuffer et al. (2019) | 22 | 40 | | | ✓ | | ✓ | | ✓ | | | | | |
| Kupin et al. (2019) | 95 | 92.86 | | | | | ✓ | | | | | | | |
| Ajit et al. (2019) | 33 | 93.03 | | | ✓ | ✓ | ✓ | ✓ | | | | | | |
| Liebers et al. (2021a) | 16 | 100 | | | ✓ | | ✓ | | | ✓ | | | | |
| Miller et al. (2020b) | 46 | 85 | | | ✓ | ✓ | ✓ | ✓ | | | ✓ | | | |
| Miller et al. (2021) | 46 | 98.53 | | | ✓ | ✓ | ✓ | ✓ | | | ✓ | | | |
| Li et al. (2024a) | 46 | | 0.057 | | ✓ | ✓ | ✓ | ✓ | | | ✓ | | | |
| Shen et al. (2019) | 20 | 95 | | | | | | | | | | ✓ | | |
| Wang et al. (2023) | 30 | 97.5 | | | | | | | | | | | ✓ | |
| Ritola et al. (2023) | 10 | 87.87 | | | | | | | | | | | | ✓ |
| Deng et al. (2023) | 25 | | | 98.1 | | | | | | | | | | ✓ |
| Suzuki et al. (2023) | 29 | 93.8 | | | | | | | | | | | | ✓ |

It is important to note that most of these approaches have been tested with different datasets and/or under different conditions. This makes it difficult to compare the relative performance of each approach *video streaming* (Wu et al. 2017). Both datasets provide three-dimensional information; additionally, the latter provides quaternion information. The results of the experiments show that the PART and LMT classifiers achieved an accuracy of more than 99% for the first dataset. On the second dataset, every classifier achieved an accuracy of more than 99%, suggesting that quaternion data can significantly simplify binary classification tasks.

Similarly to Miller et al. (2020a), Mustafa et al. (2018) found that body movement patterns (freely interacting within a VR environment) exhibit enough information to identify users. In their work, the authors argued that data of head movements can be used for authentication purposes. To prove their claim, they designed an experiment with 23 subjects that interacted with a VR application developed by the authors. The VR application was designed specifically for evaluating head and body biometrics. In the experiment, the users start from a point A and must move towards a random point B where a ball appears, then to a third random point C, and so on, until the user has reached 25 balls. To move towards a ball, users must move a VR pointer towards the target ball by nudging the VR headset in the desired direction. Each subject was evaluated in two sessions that took place, at least one day apart from each other, and the experiments showed an EER of around 7%, in the best case. Although these results would not suffice to deploy a real-life application, head and body movements hold promise for authentication purposes.

Beyond the raw data of its sensors, VR equipment can provide more interesting information for authentication purposes. In 2019, Pfeuffer et al. (2019) investigated the use of body motion and behavioural biometrics for identifying users in a VR setting. They conducted their experiments using an HTC Vive (composed of an HMD, two controllers, and an optical tracking), enriched with an additional eye tracker from Pupil Labs. First, they identified four basic tasks that can be found in most VR applications: pointing, grasping, walking, and typing. Then, they designed specific tasks that revolve around these four activities and invited 22 participants to perform them. Based on the collected data, the authors evaluated different motions and distance relations (e.g., the distance between the headset and the joysticks). However, their best result, obtained using a combination of head motion and the distance between every device, achieved only an average of 40% accuracy. Therefore, further investigations were warranted.

Other works investigated the combination of data from controllers and HMDs for achieving authentication on VR. Several of these works used a virtual ball-throwing task as a case study. In the first instance, Kupin et al. (2019) used as a feature the position of the dominant hand holding a VR controller, following the intuition that users exhibit identifiable trajectories due to their unique biomechanics. In their experiment, the authors instructed users to pick up a virtual ball and throw it at a target, and they compared the trajectory of the dominating hand of each user against a dataset containing the movements of the other users. Using bounding box centring and a nearest-neighbour (NN) matching algorithm on 95 different 3D trajectories, the authentication accuracy reached 92.86%, proving that trajectories exhibit enough peculiarities to identify a user.

Ajit et al. (2019) improved the approach proposed by Kupin et al. (2019), extending the data analysis to the recessive hand and the head and using the device orientation in addition to the position. In an experiment involving 33 users, the authors improved the results of Kupin et al., obtaining an authentication accuracy of 93.03%. Their approach is based on a perceptron classifier which learns weights on the matches between position and orientation features on the hand controllers' trajectories, as well as the headset trajectory. In practical terms, trajectories belonging to the same user result in a low classifier score and allow the user to authenticate.

Liebers et al. (2021a) further investigated the effect of body normalisation on user identification in VR. In particular, they analysed the impact of body height normalisation, arm length normalisation, and the combination of both against scenarios where no normalisation is applied. They evaluated a small pool of 16 participants during the execution of two tasks that can be correlated to actions executed in a typical VR game: a bowling throw task and an arrow shooting task. The research highlights that, while physiological factors play a clear role in behavioural biometrics, they also increase the noise injected into deep learning algorithms. Removing or reducing the features showed a positive impact on the identification accuracy of the deep learning models implemented by the authors, i.e. recurrent neural network (RNN) and multi-layer perceptron (MLP). Not only this work suggests that it is possible to achieve user identification without processing some sensitive information, but also that it is feasible to perform continuous authentication during regular gaming sessions.

Miller et al. (2020b) noted that authentication approaches based on behavioural biometrics assume that users enrol in the same VR system they will later use. However, in the future, users might interact with many different VR headsets in their daily tasks. This exposes a problem, because different VRs exhibit system-specific biases, meaning that every time that users want to use a new headset, they have to engage in new enrolment phases. Therefore, Miller et al. investigated the differences between 3 different VR systems (Oculus Quest, HTC Vive, and HTC Vive Cosmos), collecting the biometric data of 46 users performing a simple task on each system. They instructed the participants to perform a simple (virtual) ball-throwing task on each headset, repeated 10 times, for several consecutive days. Expanding on what has previously been done by Ajit et al. (2019), they introduced trigger control information from the dominant controller, as well as linear and angular velocity to account for the differences in speed between users. The results in terms of within-system and cross-system accuracy showed that some features play a prominent role for authentication purposes. In particular, head orientation, left controller position, right controller orientation, and right controller position proved to be the most important features for authentication purposes. Their results confirm the intuition of Ajit et al. (2019), about the relative importance of different features when performing authentication.

Improving their previous work, Miller et al. (2021) proposed using Siamese neural networks to learn the baseline distance function across HMDs. Knowing this information, when a user wants to use a new VR headset, it suffices to apply the distance function to offset the inherent differences with the original headset used for enrolment. This prevents users from having to go through an additional enrolment phase every time they use new VR equipment. In this paper, they used the same experimental setting they used in their previous study (Miller et al. 2020b). Regarding user authentication, their results show that the Siamese network achieves at its best a 1.39% EER, when the Oculus Quest trajectories are used at enrolment time and the HTC Vive at use time. For user identification, their highest accuracy is 98.53%, with the same setup observed for the EER authentication result.

Using the same dataset provided by Miller et al. (2020b), Li et al. (2024a) investigated whether forecasting the users' motion could allow for achieving more reliable behavioural biometrics. The authors trained a transformer-based model and verified the performance of using only the initial trajectory, versus the performance of the model when the data is enriched with the forecasted trajectory. The experiment they conducted show, first of all, that the transformer-based learning achieves on average slightly lower EERs than fully convolutional neural networks (FCNs). Besides, their experiments highlight that the transformer-based model with forecasted trajectories outperforms the same model without forecasting involved, regardless of the time window size.

Something worth noting, is that most of the works covered in this section investigate user authentication over short timescales (i.e., seconds to minutes). This leaves open questions regarding the consistency of users' movements (hence authentication success rates) over medium and long timespans, where medium timescales indicate days and long ones refer to weeks to months. Using the datasets of Miller et al. (2020b, 2022b) and Ajit et al. (2019) investigated the impact of time elapsed on authentication success rates, when using users' movements as authentication data. The authors' data analysis shows that short and medium timescales have minimal impact on VR users' behaviour, and authentication success rates do not vary. On the contrary, analysis over long timespans (7–18 months) show that users exhibit varying degrees of fluctuations in their biomechanics behaviour, which in turn impacts negatively the authentication success rates. Even though the FAR could be reduced to take into account behavioural drift, this could increase the FRR and reduce usability, leading to users' frustration. Therefore, the authors suggest that authentication routines based on biomechanics should take into account behavioural drifts and train authentication mechanisms over data that include longer separation of months to years. The paper also suggests that future work should investigate the effect of temporary conditions on the collected authentication data, for example temporary injuries that forcefully change users' movements, as well as events that affect the short timescales, such an unexpected and distracting phone call.

Shen et al. (2019) note that VR systems often store users' private data, such as social network accounts and credit card information. Although convenient, this method of accessing information exposes significant security and privacy issues. For amending this issue, the authors proposed an authentication system based on the users' gait signatures, named GaitLock, that implements a new gait recognition model, Dynamic-SRC. Dynamic-SRC fuses data information from both the HMD accelerometer and the HMD gyroscope to address the problem of noisy signals coming from head motions. The validation experiments consisted of different walking tasks; among these tasks, the most interesting is the uncontrolled outdoor walking task. The results show that Dynamic-SRC outperforms by 20% the gait recognition accuracy of the best alternative algorithm (SRC with Sparse Fusion). Furthermore, in the uncontrolled walking task, GaitLock achieves an accuracy of 95% using only a small dataset of 5 steps. Although the authors used only an AR device that could provide enough computational power (i.e., Google Glasses), they noted that with enough computational power, their system can be transposed to VR.

Wang et al. (2023) noted that VR users tend to adjust the strap to wear their VR headset tight on their head. This forms a rigid body composed of the head and the device, with relatively low deformation and constant distance between any two chosen points. Noting that the embedded speakers and microphone create an acoustic channel that would be partially absorbed and reflected by the user's skull, the hypothesis of the authors is that different skull would exhibit different absorption and reverberation, this could be used for identifying different users. They trained a CAE-CNN model and showed that the system works, with an identification accuracy of 97.5%. Besides, the approach is resilient against various attacks, such as human impersonation attacks, silicone and 3D printed fake heads, and replay attacks.

Ritola et al. (2023) enriched a VR headset with a front-attached LeapMotion sensor, which allows tracking the hands of the VR user and use them to naturally interact with the virtual environment. Their hypothesis was that hand gestures would exhibit peculiar characteristics from one user to the other, characteristics that would allow for a continuous authentication system for VR operators. In their work, they tested different combinations of tasks, sensors (i.e., head, eye-tracking, and hand data), and ML classifying algorithms. The experiments showed that, when tested on tasks similar to the training tasks, the system an accuracy of up to 99.86%, using all the aforementioned features. It is worth noting that, while the eye data was the most relevant feature, the hand data itself managed to achieve an accuracy of up to 87.87%, suggesting that hand data itself can be sufficient for a reasonable authentication system.

On the use of the LeapMotion tracking device in VR systems, Deng et al. (2023) noted that similar sensors can be used to mitigate the threat of replay attacks executed with fake robotic hands. In authentication tasks, such as an in-air signature, a robotic arm can imitate a human hand-motion, potentially leading to vulnerabilities to robotic replay attacks. The authors show that this vulnerability is easy to trigger if the signature recognition system tracks only one point, as done in the previous state-of-the-art. Instead, the authors track multiple points (i.e., 21) at different hand joints and deploy a CNN architecture for recognizing the different in-air signature characteristics of 25 participants. Experiments show that the authentication system achieves a remarkable F1-score of 98.1% and, the higher the enrolled signatures, the better the performance. To verify the resilience against robot-impersonation attacks, a robotic arm is equipped with a 3D-printed hand of the targeted user and is programmed to perfectly imitate the trajectory of the victim's hand: while other state-of-the-art signature methods exhibit a false acceptance rate (FAR) of 32.2%, their approach keeps the FAR at 0.2%.

Suzuki et al. (2023) also investigated the use of hand gestures for VR authentication purposes. PinchKey is an authentication approach that utilizes the hand tracking sensor integrated in Meta Quest 2 VR headset, which allows tracking hands' orientation and movements. To authenticate the users, the approach leverages the pinch motion executed by closing together thumb and index finger, a motion that is often used in many VR scenarios that rely on hand tracking. Experimental results conducted with 29 participants showed an accuracy of 93.8% and an average authentication time of 1.65 s. Since it leverages a common gesture, PinchKey could be used for developing a continuous authentication system, seamlessly integrated with regular VR applications. Besides, usability studies conducted on 12 participants showed that PinchKey rated good scores in the three elements of usability for human–machine interaction drawn by the standard ISO 9241-11 (i.e., effectiveness, efficiency, and satisfaction).

One thing to highlight, which applies to biometrics data in general (including eye-tracking data, covered in the next subsection), is that most studies have a limited number of study participants, as discussed by Miller et al. (2022a). This is mainly due to the costs of the equipment and the necessity of conducting in-lab studies for ensuring control over confounding variables. The limited number of samples in biometrics dataset complicates the applicability of some techniques, deep learning approaches above all. For this reason, Miller et al. (2022a) propose two forms of data pre-processing for improving deep neural networks (DNNs) performance: modelling of spatial relationships between headset and controllers, and input trajectory smoothing to remove tracking noise. Although promising on paper, results were limited in magnitude (i.e., around a 2% improvement), warranting further studies.

## 5.3 Authentication with eye-tracking sensors

Although early VR headsets were not equipped by default with eye trackers, these sensors are becoming more and more common in commercial headsets. One of the main benefits provided by VR HMDs to eye tracking is that HMDs fully cover users' eyes, allowing for controlling lighting conditions and reducing problematic reflexions.

Authentication using eye tracking sensors has been studied long before VR technology took hold. For example, in 1978, Hill (1978) patented an apparatus for recognising individuals by the retinal vasculature in their eyes and, in 2004, Maeder et al. (2004) investigated the use of eye gaze to authenticate users. More recently, Sluganovic et al. (2016) observed that gaze-based authentication systems suffer from high error rates or require long authentication times. Based on this premise, they proposed using reflexive saccades, rapid eye movements that reorient the eye to the next focus object. Other works, such as Rigas et al. (2016) and Holland and Komogortsev (2013), show the potential of using saccades for identification purposes. In this section, we focus on works that discuss the use of eye-tracking sensors in VR, either alone or alongside other sensors.

For the scope of this paper, we neglect the research conducted in the vast field of eye-based authentication and recognition, not specifically targeted to VR applications and implementation. In Table 4, we summarise the papers discussed in this section, and we highlight which features have been used by the authors for performing authentication.

**Table 4** Works that have analysed authentication and identification with eye-tracking sensors

| | Participants | Accuracy (%) | Equal error rate (%) | F1-score (%) | Head position | Head orientation | Controllers position | Controllers orientation | Gaze | Saccades | Pupil size | EOG signals | Blinking |
|---|---|---|---|---|---|---|---|---|---|---|---|---|---|
| Rogers et al. (2015) | 20 | 94 | | | ✓ | ✓ | | | | | | | ✓ |
| Liebers et al. (2021b) | 11 | 100 | | | ✓ | ✓ | | | | ✓ | | | |
| Luo et al. (2020) | 70 | | 4.97 | | | | | | | ✓ | | ✓ | ✓ |
| Khamis et al. (2018) | 26 | 82 | | | | | | | ✓ | | | | |
| Lohr et al. (2018, 2020) | 458 | | 9.98 | | | | | | ✓ | ✓ | | | |
| Olade et al. (2020a) | 15 | 98.6 | | | ✓ | ✓ | ✓ | ✓ | ✓ | | | | |
| Tricomi et al. (2023) | 34 | | | 90 | ✓ | ✓ | ✓ | ✓ | | | ✓ | | ✓ |
| Zhu et al. (2020) | 52 | | 4 | | | | | | | | ✓ | | ✓ |
| LaRubbio et al. (2022) | 10 | 75 | | | | | | | ✓ | | | | |
| Lohr et al. (2023) | 5 | | 20 | | | | | | ✓ | ✓ | | | |
| Ritola et al. (2023) | 10 | 99.62 | | | ✓ | ✓ | | | ✓ | | ✓ | | |

It is important to note that most of these approaches have been tested with different datasets and/or under different conditions. This makes it difficult to compare the relative performance of each approach

Rogers et al. (2015) showed the possibility of identifying users using their unconscious head movement and blinking, while performing a simple passive task. The authors set up an experiment in which they showed 20 participants a rapid series of numbers and letters. Using the collected data, they achieved an identification accuracy of 94%, showing the suitability of unconscious movements for unobtrusive user authentication.

Liebers et al. (2021b) proposed an implicit identification method based on users' saccadic movements and head orientation. Similarly to when eyes switch from slow pursuit to saccadic movement, as previously mentioned, when the stimulus speed exceeds the saccadic speed threshold, human beings must recruit head movement to keep track of the stimulus. The authors show that the characteristics of fixations and saccades, together with head orientation, are complex and unique enough to be used as biometric features. In particular, with a small pool of 12 participants, using a KNN algorithm on every feature they envisioned, their approach achieves an accuracy of 82%, whereas using deep learning the accuracy reaches the 100% mark.

Luo et al. (2020) noted that authentication methods based on head and body movements, such as those mentioned above, expose the authentication actions to bystanders. They also noted that eye-gazing alone can exhibit a high EER, too high for trusting it as an authentication method. Therefore, with OcuLock the authors propose to involve the entire human visual system (HVS), as it is composed of different entities, such as eyelids and extraocular muscles, which exhibit features that could be used for authenticating purposes. Tested with 70 participants, OcuLock achieves an EER of 3.55% against impersonation attacks and 4.97% against statistical attacks. However, it should be noted that standard HMS headsets equipped with eye-tracking cameras cannot measure such HVS signals; for example, they cannot be used to implement a framework based on electrooculography (EOG) such as OcuLock. In fact, to capture EOG signals, the authors had to enrich with tiny electrodes the foam face cover of a Lenovo Mirage Solo VR headset.

With VRPursuits, Khamis et al. (2018) showed that VR equipped with eye tracking is suitable for implementing smooth pursuits that address the Midas touch problem (i.e., the problem of distinguishing deliberate gaze from basic eye perception functions). Although their work does not address user authentication nor security properties, their implementation of a virtual ATM shows the potential feasibility of an authentication method based on smooth pursuits detection. In their proof of concept, participants were asked to enter a 4-digit PIN by intentionally looking at the desired numbers. The numbers, from 0 to 9, were visualised as two groups of cubes rotating clockwise and counterclockwise, respectively.

Lohr et al. (2018, 2020) explicitly address the goal of using eye movement biometrics to authenticate VR users. During the enrolment phase, the prototype extracts four fixation features and eight saccades features, for a total of twelve eye movement features. To perform authentication, the authors propose to compare the features of two different templates (the user currently examined and a stored template), create a vector of match scores, and finally combine the scores into a single fused score using a weighted mean fusion with equal weights.

Olade et al. (2020a) proposed BioMove, a method for recognising biomechanics behaviour patterns and using them to authenticate users. In particular, the authors studied the movements of the head, hands, and eyes of 15 users while performing controlled tasks, such as grab, rotate, and drop. The authors investigated the use of BioMove for the purpose of both user identification and authorization. In the best case scenario, for identification tasks, BioMove achieved a classification accuracy of 98.6%. Regarding authentication, the authors' experiments show a 0.0032% FPR and a 1.3% FNR, highlighting that BioMove (and behavioural biometrics, in general) could be used as an efficient second-factor authentication.

Tricomi et al. (2023) have done a similar investigation to the one conducted by Olade et al. (2020a), although using the eye-blinking as a feature, instead of the gaze. Their results confirm what has been previously shown in the state-of-the-art, showing that users can be identified based on the biometrics data collected during their interaction in a VR environment (with an F1-score of approximately 90%), and that eye-tracking data is the sensors that provides the most relevant information for identification goals.

Zhu et al. (2020) noted that, by design, VR systems prevent their users from seeing what happens around them in the real world. This means that users have a hard time concealing their actions from bystanders, which could lead them to potentially leak sensitive information. BlinKey, an authentication scheme designed for VR headsets equipped with eye-tracking technology, aims to mitigate the problem by using only users' eyes, naturally covered when using a VR HMD. The core idea of BlinKey is to encode a password as a set of blinks, performed with a rhythm only known to the user (e.g., following a tune that they like). As an additional authentication layer, BlinKey analyses the pupil size variation between blinks, a biological marker that uniquely characterises every person. BlinKey was implemented and tested using a commercially available HTC Vive Pro equipped with a Pupil Labs eye tracker. To evaluate the security resilience of their proposals, the authors tested four different types of attack: zero-effort, statistical, shoulder surfing, and credential-aware. For the first attack, attacking participants tried to guess the target blinking pattern without any information and were incapable of compromising passwords composed of 7 or more blinks. With statistical attacks, attackers had access to a dataset of blinks of their victim, which they could statistically analyse. This leads to a slight advantage for the attacker, with respect to the zero-effort attack, but passwords composed of at least 8 blinks seem to prevent any compromise attempt. Shoulder-surfing attacks were more successful, with an attack success rate of 4.9%. Last, credential-aware attacks assume that the attacker knows the password and tries to replicate it with their own blinks (i.e., exhibiting their own pupillary biometrics). Although the attack proved to be more successful than the previous ones, when the password consisted of at least 10 blinks the attack success rate was 4.4%, showing that pupil variation alone cannot be used as an authentication factor.

As previously mentioned, Ritola et al. (2023) tested the performance of various ML algorithms on different sensors data and their combinations (i.e., head, eye-tracking, and hand data) for delivering continuous authentication in VR. Regarding the eye-tracking data, they showed that a latent Dirichlet allocation (LDA) model, achieves an accuracy of up to 99.62% when tested only eye-tracking data and testing tasks similar to the training ones. When tested on different tasks, the system still achieved a good performance of 88.57%, again only using eye-tracking data.

In their short paper, LaRubbio et al. (2022), after noticing that the literature regarding gaze-based authentication in VR focused on image viewing tasks, hypothesised that similar performance could be achieved with two different tasks: viewing a random positioned dot, and completing a complex VR simulation (in their case, a nuclear reactor startup procedure. While their experiments with 10 participants confirmed the state-of-the-art conclusions, with the image viewing task achieving around 75% accuracy, the two other scenarios proved to be insufficient. The random dot viewing achieved only a 15% accuracy and the complex scenario a mere 12%. Although more in-depth experiments should be done, the authors concluded that integrating image-viewing scenes during the training program could be a viable solution for enabling within-task authentication. Another short paper from Lohr et al. (2023) shows that it is feasible to train a CNN model over an intentionally degraded eye-tracking dataset, deploy it on a system with different users that utilise a different VR technology, and achieve a better performance than chance (i.e., 20% EER).

### 5.4 Other works in VR authentication

Most papers in the field of VR authentication, at the time of writing, either leverage something users know, such as a password or a set of secret interactions, or some personal characteristics, such as their behavioural biometrics.

For example, something that is common to all the papers covered in Sect. 5.2 which base their authentication strategy on users' movements, is that trajectories could be in principle observed and replicated. Besides what has been demonstrated by Gopal et al. (2023), previously mentioned in Sect. 3.3, in a short paper Miller et al. (2022c) pitch an impersonation attack that leverages 2D video recordings of VR users performing authentication tasks to reconstruct their 3D trajectories. Preliminary experiments show how, in a best-case scenario, an attacker could achieve an attack success rate up to 82.2% and an EER as low as 11.4%.

Combining different approaches for achieving multi-layered authentication has the potential to offset similar threats, and produce stronger authentication schemes across the board.

With RubikBiom, Mathis et al. (2020a) investigated the use of knowledge-driven behavioural authentication in the VR domain. The authors developed a proof of concept that uses two different authentication features. A password only known to the user and the movement patterns that arise from the act of inputting such a password in VR. As for previous proposals for behavioural biometrics, the intuition is that every user exhibits peculiar movement patterns due to their specific biomechanics. In particular, an attacker that attempts to infiltrate a victim's account by shoulder-surfing, not only should steal the password, but should also be able to replicate the same movement patterns when inserting the password. In RubikBiom, the password is a 4-digit PIN selected on a 5-faced cube. Each face has a different colour and exhibits 9 digits (1–9), similarly to the setup used by the same authors in their works previously discussed (Mathis et al. 2020, 2021b). The best accuracy score was 98.91%, which demonstrates that multilayered approaches in VR hold promise. The result was obtained with a fully convolutional network (FCN) over a multivariate dataset (composed of dominant hand position and rotation), together with non-dominant hand position and rotation.

In a preliminary study, Wang and Gao (2021) show that multi-attribute authentication methods could also be effective in countering Man-in-the-Room (MITR) attacks. The basic intuition behind their authentication scheme is that objects in the real world are defined by a rich set of features (e.g., colour and shape). In their proof-of-concept, the user chooses a number of secret features at enrolment time and, at authentication time, the framework creates random objects which might contain none, some, or every feature. To log into the system, the user must stare at the object that exhibits all the desired features. This approach is particularly interesting because it allows to effectively decouple the input process from the password, increasing the resilience to MITR attacks.

von Willich et al. (2019) discussed the danger that passersby create to VR users, unaware of their surroundings. HMDs impede visual perception of the surrounding environment, allowing bystanders to physically collide with VR users or observe their actions to infer their authentication credentials. As a mitigation, the authors propose to detect passersby using environmental sensors and depict them in the VR environment, so that the VR user is aware of them, effectively implementing augmented virtuality (AV), a paradigm that involves enriching the virtual world with information from the real world. This depiction can take different forms, such as avatars, 2D pictures, or 3D renderings. While AV was originally meant for avoiding unwanted collisions with dynamic agents in the real world, the authors argue that it can be used for preventing some by-standing attacks. Although AV cannot impede attackers from observing from a distance, it could be used to detect attackers that get close enough to be picked by the sensors and prevent them from having a proper close-up view of VR users' motion, thus reducing their chances of inferring useful information.

Some works in the literature discuss the use of additional sensors for authenticating VR users. Progonov et al. (2023) proposed a passive acoustic sensing approach, with an ad-hoc-device that allows for sensing the bioacoustic signals produced by the human body movements (specifically, wrist joints and cervical vertebrae noises). After modifying two devices, a VR headset and a smartwatch, with a high-sensitivity electret condenser microphone, the authors conducted experiments with a limited pool of 3 participants. The authors analysed the collected data with two different classifiers (i.e., RF and MLP) and compared their results with the false acceptance rate (FAR) and false rejection rate (FRR) of other state-of-the-art methods. The experiments show that the approach is viable, in particular when paired with the MLP.

Cha et al. (2022) noted that, while speech recognition could be used for implicit authentication on VR, noisy environments would render sound-based recognition approaches complicated. They propose to use a silent speech recognition (SSR) method, an approach that can be compared to lip-reading, usually achieved with sensors attached around the user's lips. The authors highlight that the traditional sensors positioned on the lips would require VR users to utilise an additional wearable device. Therefore, in their paper they integrate a set of facial electromyograms (fEMGs) in the VR visor, making so that the sensors would connect to the area around the user's eyes. By using a DNN trained on 20 participants' data, the experiments show that the system achieves an accuracy of 92.53% when faced with the task of classifying six silently spoken words.

Li et al. (2023) propose SigA, an authentication process based on the use of remote photoplethysmogram (rPPG), a method that relies on the use of cameras pointed at the user's face to capture changes in skin colour and blood flow. In SigA, the authors claim that it would be feasible to install a face-oriented microcamera inside a VR headset, and use the subtle signals of the user's skin to authenticate them. To verify the feasibility of the method, they trained a classifier model on a set of publicly available rPPG datasets and show that the approach can achieve noteworthy results if employed for continuous authentication, with an F1-score of over 95% when using a video sample of only 1 s. The authors also showed that the system is quite resilient against impersonation and statistical attacks.

Grandi et al. (2023) pitch the use of a small device equipped with a haptic motor and a button, mounted on the side of a VR headset. Through the motor vibrations, the device generates a 4-digits one-time-password (as a sort of Morse code) and the user must replicate the dots and dashes signals by pushing the integrated button. Intuitively, it is difficult for a by-stander to perceive or hear the motor vibrations, hence the method should be resilient against observation attacks, but to date the method has not been tested in any controlled experiment.

## 6 VR for teaching, training, and evaluating security and privacy

VR environments have been proposed as a mean for improving the cyber and physical security of sensitive locations, as well as for training employees on routines and policies. In addition, VR has been proposed to assess the usability of authentication processes.

In this section, we aim to answer to RQ4, previously presented in Sect. 1.1, providing an overview of the different security applications that could benefit from the use of VR. In Table 5 we summarise all the works we cover in this section.

**Table 5** Works on the use of VR for teaching, training, and evaluating cybersecurity and privacy

| Authentication approach | Works in the Literature |
|---|---|
| Teaching cyber-security | Puttawong et al. (2017), Visoottiviseth et al. (2018), Veneruso et al. (2020), Ulsamer et al. (2021), Chekhovskoy et al. (2022), Munsinger et al. (2023) |
| Teaching privacy | Lim et al. (2022), Roskó et al. (2022) |
| Training physical and cyber-physical security | Henrique et al. (2015), Chandrashekar et al. (2023) |
| Evaluating cyber-security usability | Mathis et al. (2021a), Mathis et al. (2022c), Mathis et al. (2022a), Mathis et al. (2022b), Li et al. (2024b) |

### 6.1 Teaching cyber-security

In 2017, Puttawong et al. (2017) noted that the abstract nature of network security can make it hard to grasp for students, when taught in traditional lecture-based classes. Therefore, the authors proposed VRFiWall, a VR edutainment game developed with Unity for teaching firewall security. In this game, users interact with the environment using two basic actions, that is, pointing and head gestures. This simple game, which instructs its users about stateless firewalls, revolves around a hero on a secret mission from his kingdom (IP source) to another kingdom (IP destination). To complete his mission, the hero has to match the requirements of Statelessa, a non-playing character that embodies the firewall.

Following the same philosophy, Visoottiviseth et al. (2018) proposed Lord of Secure, a serious game for VR that challenges users with different cybersecurity concepts. The game is divided into three chapters, which present various topics: IP spoofing, flooding, TCP covert channels, firewalls, intrusion detection and prevention systems (IDSes and IPSes), and honeypots. The goal of Lord of Secure is to make abstract security topics easier to digest for cybersecurity students, and to evaluate the users' understanding of core concepts by means of pre-test and post-test quizzes. Among a group of 28 participants, 90% claimed that they well understood the concepts covered in the game, and 82% thought that this approach allowed them to understand better than traditional frontal lessons. Indeed, quizzes that included pre-test and post-test clearly showed that the users improved their knowledge by playing the game. On average, considering Chapters 1 and 3, 70% of the users improved their scores after playing the game. Taking into account users' enjoyment and their improvements, the authors concluded that edutainment games might be an alternative and more effective way to teach cybersecurity.

Veneruso et al. (2020) developed a VR game that aims to educate users on cybersecurity topics, named CyberVR. For CyberVR, the authors equipped the headset with a LeapMotion hand-tracking sensor that allows a more natural user interaction than VR controllers. This serious game is split in 6 mini-games that require the user to complete various tasks for ensuring that the initial insecure IT system becomes secure. Various cybersecurity related concepts are taught in the game: network mapping (NMAP), illegal information flows, code injection, patch management, dynamic software analysis, privilege escalation, and public-key cryptography. A user study, conducted with 40 participants, show that CyberVR is equally effective than traditional learning methods, but more engaging for the students. The authors note that the novelty of the equipment could impact the sustainability of these outcomes, either negatively or positively, and point to the need for further longitudinal studies.

Recently, Ulsamer et al. (2021) discussed the use of storytelling in VR to improve user awareness of information security (ISA), in particular, with respect to social engineering. In their experiments, they split the participants into two groups, providing one of the groups with a VR environment and the other (named no-VR) with a traditional e-learning platform. The learning objectives set for the two groups were the same. The no-VR group received theoretical material and interesting examples (but no videos) related to social engineering. The VR group, instead, watched a storytelling 3D video with a coherent and immersive plot that revolves around a hacker which performs social engineering; the hacker is played by an actor and the user follows his steps throughout the story. The ISA analysis show that the VR group achieved better scores in the ISA test, supporting the hypothesis that VR-based learning can improve cybersecurity awareness and material retention. Since that the no-VR group was provided with static material but no traditional videos, it is still unknown whether regular videos could achieve comparable results to VR-based videos.

Also Chekhovskoy et al. (2022) investigated the use of VR for training information security specialists. First, the authors selected a set of 7 network attacks with a relatively high complexity of visualisation. Then, they implemented in Unity an environment that represents through fantasy-inspired objects (a castle, a river, and so on) various network parts, tools, and technologies. Besides having to answer theoretical questions, trainees experience the attacker point of view by moving in the virtual world and attempting to infiltrate the castle and steal a treasure. After this, users train their defending capabilities; in this case, the character starts from inside the castle and trainees must move around and physically patch the various vulnerable accesses. As a last step, the trainees get to attack the castle again, such that they can experience first-hand the effectiveness of the defences they implemented in the previous phase. A pool of 28 participants was split in an intervention and a control subgroup, and the experiment show a higher material assimilation rate in the subgroup using the VR (i.e., 86%), versus the subgroup that were not offered the VR experience.

Munsinger et al. (2023) focus on security operation centres (SOCs), the set of processes and professionals within an organisation that prevents, monitors, and responds to cybersecurity threats and incidents. As noted by the authors, considering the volume and variety that organisations have in terms of data and networking infrastructure, SOCs tasks are complex. Incident event management, data fusion, and anomaly detection are helpful but not sufficient. It is necessary to train the cyber-situational awareness of SOC operators. The authors developed a VR-based network and compared the performance of operators versus the training performed with traditional tools. The 12 participants to the study were assigned the task of analysing the same multi-stage network attack packet capture (PCAP), either using the network analysis tool develop in VR, or the traditional Wireshark, or both. Albeit the participants pool is small, VR-based tools show promise in enhancing SOC operators' awareness, particularly when paired with traditional tools and when proposed to novice operators.

### 6.2 Teaching privacy

Lim et al. (2022) created a VR role-playing game for increasing privacy awareness in users. Split in two sessions, first data collection and then data utilisation and distribution, the role-playing privacy tutorial allows users to control an avatar that moves coloured spheres from a source to different destinations, as a metaphor of data elaboration and communication channels. The spheres represent different personal data (e.g., physical features, voice interaction, or movement data); the points of interest are the VR user producing the data, data centres, content providers, and third parties that might have access to private information. The avatar can select categories of personal data, visualise it and analyse it, and move it to the different destinations that such information could reach in real case scenarios. Quantitative and qualitative results of the study, conducted with 14 participants, show that the role-playing game achieves the same scores of conventional text-based privacy policies, but achieves higher satisfaction and usability scores.

Roskó et al. (2022) conducted a survey among university students to verify their baseline knowledge within the topic of privacy and privacy awareness. The proposed questionnaire included correct answers, random wrong answers, and well-known (often picked) wrong answers. For some questions proposed, a considerable number of students (up to 36.67%) gravitated towards the well-known wrong answers, highlighting the need for intervention. The authors designed a virtual environment that included descriptions of case studies, videos, interactive tools for checking passwords, forums for interacting with other VR users and exchange information. Besides, the chosen platform (MaxWhere) allows for designing realistic environments, including organised IT workspaces and related tools. Unfortunately, the paper does not present relevant data that could be used for assessing whether this intervention had a positive impact on students' privacy awareness, or not.

### 6.3 Training physical and cyber-physical security

Henrique et al. (2015) proposed a VR-based tool to train staff in a nuclear facility structure and plan its defence strategies. The authors produced (and used) a high-fidelity 3D reproduction of a Brazilian nuclear research centre, showing that their approach can be useful for evaluating the physical security of different facilities. In particular, they prove that inside the VR simulation, it is possible to identify strategic points of view and change the environment to evaluate the impact on visibility. For example, their VR implementation allows them to change weather conditions, such as rain, wind, and snow, as well as natural light conditions (day and night alternation, sun position, presence of stars) and artificial illumination. In conclusion, VR environments can be useful tools, both for evaluating the physical security of critical facilities and for training employees in risk management.

Chandrashekar et al. (2023) noted that critical infrastructures are transitioning more and more to cyber-physical systems (CPSes) through digitization processes. Besides all the positive implications arising from this evolution, there are also new needs that must be addresses. Above all, the need for innovative cyber-physical security training revolving around these critical infrastructures. In their paper, the authors propose an IoT-based system that simulates a wastewater treatment facility, paired with a VR environment. The combination of technologies allows for creating a complete digital-twin immersive platform, in which employees can experience the results that cyberattacks and intrusions can have on the physical equipment. With this training, employees learn to recognise cyberattacks by effects and side effects they have on the equipment, and respond efficiently to the threats.

### 6.4 Evaluating cyber-security usability

Mathis et al. (2021a) and Mathis (2021) noted that evaluating the usability of new authentication schemes can be expensive, in particular for the cases when special hardware and infrastructures are required. VR can be helpful in addressing this problem. First, it saves researchers time and money by sparing them from building their physical prototypes (e.g., a replica of an ATM). Moreover, it enables them to recruit more participants and increase their pools' diversity, by allowing to enrol people from different parts of the world.

With RepliCueAuth, Mathis et al. (2021a) paved the way to usability studies in VR, discerning which security assessments could be transferred from VR to the real world and which ones not. In their work, the authors replicated in VR the experiments of a paper that explored 3 cue-based authentication methods on situated displays. The results showed that some usability studies can be transferred from VR to the real world, such as the accuracy of authentication entry, perceived workload, and perception of security. However, there are some notable differences between the VR experience and the physical study, in terms of results. Users took longer time to authenticate on VR with the touch-based method, while they in the real world the gazing-based approach was the slowest one. Similarly, the security studies conducted on VR and on real world experiments lead to comparable results for some evaluation variables, but results differed in other instances. For example, the attack rates for shoulder-surfing attacks were similar when observing a real human being and a VR avatar, but the accuracy of the attackers' guesses was different.

In another study, Mathis et al. (2022c) further investigated the challenge of evaluating *in situ* new authentication systems. They highlight how evaluations of new physical authentication systems (e.g., on ATMs) are often undertaken in traditional lab environments, with a lack of immersion that poses questions on the ecological validity. Building realistic replicas of some machines, such as ATMs, can have relevant costs in terms, therefore the authors sought to verify whether this could be replaced with a realistic VR replica. For their quantitative and qualitative experiments, they built an ATM and then a VR environment. Results showed that experiments on VR took less time than with a real ATM placed inside the lab, and the overall sense of immersion was the highest when using a realistic VR environment.

On a similar note, Mathis et al. (2022a) investigated also the use of VR environments for studying more efficiently shoulder-surfing attacks. They conducted different experiments where the 18 participants could observe a virtual avatar inputting sensitive information on different displays (ATMs and smartphones) and move around the avatars to find the perfect angle for spying on the targets. Compared to non-immersive scenarios, such as observation of 2D recordings, the VR-based approach showed a higher sense of being part of the authentication environment, a greater feeling of spatial presence, and a higher level of involvement. The authors also provided four recommendations to guide the decision on which scenario should be used in research, whether the 2D or the VR-based. Namely, account for real-world factors if they are of relevance (e.g., avatars reacting to proximity), proper introduction to VR shoulder surfing methods prior to data collection, keep 2D as a baseline measure as often times VR-based approaches do not outperform them, and use VR to study shoulder surfing scenarios that would be hard to pull off in real-life experiments due to constraints (e.g., ethical challenges when spying on ATMs users).

The same research group, also investigated the usability of VR for conducting remote evaluation of real-world AR authentication systems (Mathis et al. 2022b). After running experiments with 25 different participants, the authors concluded that VR can be a useful tool for involving different participants in research studies on usable security, particularly when used as a complement to traditional lab research.

Partially related to papers that fall in the category of usability evaluation, Li et al. (2024b) analysed the capability of three classifiers to detect the level of user familiarity with VR, based on the data collected during the act of opening a virtual door, locked with a PIN-keypad. Since all participants were familiar with the task of opening a PIN-locked door, the results show that it is possible to differentiate with approximately 80% accuracy whether the participants have already used VR or not (based on the variability in their movement patterns). Although the goal of the paper is focused on users' familiarity with the VR technology, hence it is not directly connected to security aspects, the fact that the authors used a security-related task is interesting, as it suggests that it might be also possible to evaluate users' familiarity with security measures and interfaces, using VR.

## 7 Conclusion

Once a technology only used in exhibitions and research laboratories, VR is becoming more and more common among local businesses and families. Therefore, it is of paramount importance to take into consideration the issues that this groundbreaking technology can create. In this work, we have provided a thorough analysis of the privacy and security threats that affect VR. First, we split the privacy issues in three main categories, explaining what could cause these issues and what might be the consequences. Then, we focused on the security aspect of VR. We categorised the threats to security in VR, including both the generic threats that affect VR and the threats that are specific to VR. We have also dedicated a specific section to the topic of authentication in VR, as it appears to be the most common area of research in the field of VR cybersecurity, at the time of writing. Finally, we have covered other interesting areas of research that use VR for cybersecurity goals, such as teaching cybersecurity, training physical security, and evaluating the usability of cybersecurity solutions.

**Funding** Open access funding provided by Örebro University. This work was partially supported by the Wallenberg AI, Autonomous Systems and Software Program - Humanities and Society (WASPHS) funded by the Marianne and Marcus Wallenberg Foundation and the Marcus and Amalia Wallenberg Foundation.

**Publisher's Note** Springer Nature remains neutral with regard to jurisdictional claims in published maps and institutional affiliations.